\documentclass[%
 reprint,
 amsmath,amssymb,
 aps,
]{revtex4-2} 

\usepackage{graphicx}
\usepackage{dcolumn}
\usepackage{bm}
\usepackage{hyperref}
\usepackage[mathlines]{lineno}

\newfam\hebfam
\font\tmp=rcjhbltx at10pt \textfont\hebfam=\tmp
\font\tmp=rcjhbltx at7pt  \scriptfont\hebfam=\tmp
\font\tmp=rcjhbltx at5pt  \scriptscriptfont\hebfam=\tmp

\edef\declfam{\ifcase\hebfam 
     0\or1\or2\or3\or4\or5\or6\or7\or8\or9\or A\or B\or C\or D\or E\or F\fi}

\mathchardef\shin   = "0\declfam 98 

\usepackage{xcolor}
\usepackage{soul}
\usepackage{mathtools}

\begin{document}

\title{Excitable dynamics and coral reef formation: A simple model of macro-scale structure development}

\author{Miguel \'Alvarez-Alegr\'ia}
\author{Pablo Moreno-Spiegelberg}
\author{Manuel A. Mat\'ias}
\author{Dami\`a Gomila}
 
\affiliation{
 Instituto de F\'isica Interdisciplinar y Sistemas Complejos, IFISC (CSIC-UIB), Campus Universitat Illes Balears, E-07122 Palma de Mallorca, Spain}


\begin{abstract}
In this work, we demonstrate that key aspects of the dynamical behavior of coral reefs at the macro scale, which evolve over time scales of centuries, can be accurately described using a model that integrates a few fundamental ecological and physical mechanisms. The model displays excitable behavior generating, among other dynamical regimes, traveling pulses and waves, which result in the formation of spatial structures resembling those observed in real reefs, without involving a \textit{classical} pattern formation mechanism, like the Turing scenario. We conduct an in-depth exploration of the bifurcations exhibited by the model as a function of the two most ecologically significant parameters. This establishes the groundwork for using the presented model as a tool to explain coral reef formation.


\end{abstract}

\maketitle


\section{\label{sec:Introduction}Introduction\protect}

Tropical coral reef ecosystems, often compared to tropical rainforests for their remarkable biodiversity, owe their existence to coral polyps that form calcium carbonate structures. These ecosystems thrive in warm, shallow oligotrophic waters.
These ecosystems are essential habitats for a wide array of marine species, with about $25\%$ of all marine life relying on them at some point in their life cycle \cite{Mulhall2009}.
Moreover, coral reefs provide natural protection to coastal regions, shielding coastlines from erosion and storm damage. 
From an economic perspective, they support local fisheries and attract tourists to various activities \cite{Fezzi2023}. 
Regrettably, coral reefs have experienced about a $50\%$ reduction in cover globally from 1957 to 2007  \cite{Eddy2021}. Furthermore, climate change poses increasing threats to
coral reefs, with global warming inducing coral bleaching \cite{Douglas2003} and ocean acidification gradually deteriorating reef structures \cite{Mollica2018,Hoegh-Guldberg2007}. Coral reefs are also jeopardized by direct human actions \cite{Halpern2008}, such as overfishing, pollution, and the introduction of invasive species, including pathogens, all of which compromise their health and biodiversity.

Coral reefs are predominantly formed by the calcium carbonate skeletons of coral polyps, marine invertebrates belonging to the subphylum Anthozoa \cite{Veron2011}. These polyps expand and reproduce both clonally and sexually while competing for resources and space with other polyps and different organisms. The limiting factors of coral growth include nutrient and oxygen availability, which are closely linked to ocean currents, water temperature, light levels, and ocean acidity. These physical variables interact in complex ways, influencing the overall condition of these critical ecosystems \cite{Wallace2011}.
Various coral species develop into large assemblages known as colonies, consisting of numerous polyps from the same species. The accumulation of multiple colonies leads to the formation of a reef. 
Coral reefs can display distinct patterns such as fringing, barrier, or atoll structures \cite{Earle2019}. A seminal study \cite{Mistr-Bercovici} used Alan Turing's theory of biological pattern formation to elucidate how coral structures consistently arise through the interaction of competing chemical reactions and diffusion processes. This Turing-like model of coral reefs \cite{Mistr-Bercovici} highlights the importance of water flow and nutrient allocation in shaping coral morphologies, with various flow regimes resulting in distinct coral structures.
Another influential study \cite{Nakamura2007} investigated the morphological evolution of coral reef topography, offering a comprehensive reproduction of the geochemical processes occurring within reef systems. The authors emphasized the importance of non-linear mechanisms in driving these processes and proposed a rationale for the diverse observed morphologies, relating them to sea level rise and diffusion effects.
Additionally, it is essential to mention a groundbreaking study \cite{Mumby2007}, which introduced a mean-field model, without space dependence, to analyze competition among corals, algal turf, and macroalgae under corals stress conditions. This approach was recently revisited \cite{Hock2024} within the framework of transient dynamics. Together, these two studies underscore the critical role of external factors, which can be represented as bifurcation parameters within the model, in affecting the stability scenarios of the ecosystem.

Here we propose a mathematical model to describe coral reef dynamics at the reef scale. A significant feature of this model is its ability to simulate coral reef formation, accurately representing their distinctive macroscopic shapes (spanning several hundred meters) based on processes occurring at smaller scales (meters), down to individual coral colonies.
We connect the genesis of these structures with spatiotemporal excitability, which is known to produce patterns such as expanding rings, target patterns, and spirals \cite{Meron1992,Mikhailov1990,Kapral1995,Alonso2016}. 
The emergence of such biological patterns from spatio-temporal excitable dynamics has previously been reported in several mathematical models of sessile species other than corals. These include vegetation models \cite{Marasco2014,Fernandez-Oto2019,Iuorio2021,Ruiz-Reynes2023} and fungi \cite{Karst2016,Davidson1997}. The spatiotemporal dynamics of these models has been tied to type-I excitable media \cite{Arinyo-i-Prats2021} and recently linked to a combination of positive and negative feedback mediated by an inhibitor \cite{Moreno-Spiegelberg2024}. 
Many such models incorporate a self-induced toxicity mechanism, leading to negative feedback. In our model, the elevation in coral reef height due to aragonite accumulation increases mortality near the surface, introducing negative feedback via aragonite. This and every other mechanism incorporated into this model are well-documented individually in the literature; however, they have not been previously considered together within a unified framework, as presented here.

The paper is structured as follows: Section \ref{sec:TheModel} introduces the model and explains the significance of each term in the equations. Section \ref{sec:Homogeneous Steady States} describes the homogeneous stationary solutions and their elemental bifurcations. Section \ref{sec:Global Bifurcations} examines the structure of global bifurcations in the system, which is essential for understanding the spatiotemporal dynamics of non-steady solutions. In Section \ref{sec:TP}, we analyze the dynamics of travelling pulses that arise due to spatial excitability. Section \ref{sec:Discussion} contextualizes the mathematical results from Section \ref{sec:TP} within the ecological framework of coral reef formation. Finally, Section \ref{sec:Conclusions} offers some concluding remarks.

\section{\label{sec:TheModel}Model\protect}
 
We present a simple mathematical model for coral reef formation, formulated as a system of partial differential equations describing the temporal evolution of two fields, \(P(x,y,t)\), representing the dimensionless density of living coral (polyps) at location \((x,y)\), and \( A(x,y,t)\), representing the density of aragonite accumulated through accretion. The aragonite field can be translated to the height of the coral reef above the sea floor via the relation:
\begin{equation}
    \label{eq:Height}
    H(x,y,t) = \frac{H_S A(x,y,t)}{\rho_A},
\end{equation}
where \( H_S \) is the sea level measured from the sea floor, and \( \rho_A \) is the bulk density of aragonite. Here we assume that both the sea level and the sea floor remain flat and constant over time, although these conditions could be relaxed in a more complete model. This assumption allows us to write any mechanism involving the reef height directly in terms of $A$, as Eq. (\ref{eq:Height}) implies. The equations governing the temporal evolution of $P$ and $A$ are given by
    \begin{widetext}
    \begin{align}
        \partial_t P &= \Big[(g_L+fP+\ell A)R - (m+sP^2+dA^2)\Big]P + g_C R|\bar{\nabla} P|^2 + g_D\nabla^2 P \label{eq:O-dtP} \\
        \partial_t A &= a(\Omega-1)P - e A \label{eq:O-dtA}
    \end{align}
    \end{widetext}
where $\bar{\nabla}=(\partial_x,\partial_y)$ and $\nabla^2=\partial_{x}^2+\partial_{y}^2$.

\( R(x,y,t) \) in Eq. (\ref{eq:O-dtP}) is a field representing a dimensionless resource indicator which measures the similarity of water composition near corals to that in the open sea. It ranges from 0 to 1, \( R = 1 \) signifying ocean-like conditions, while \( R = 0 \) indicates complete depletion of a limiting resource. Oxygen, rather than dissolved nutrients, is often the primary limiting resource in this context \cite{Nelson2019,Hughes2022}. Corals are among the most efficient organisms adapted to thrive in oligotrophic waters, where excess nutrients can lead to eutrophication, subsequently reducing coral populations \cite{Fabricius2011,Mumby2007}.
\( \Omega(x,y,t) \) in Eq. (\ref{eq:O-dtA}), represents the aragonite saturation state, calculated as the ratio of dissolved calcium and carbonate ions to the solubility product of aragonite. This field measures how easily corals can accrue aragonite.  For simplicity, both \( R\) and \( \Omega\) are considered constant in this work, although they would diffuse and be advected by ocean currents. Considering \( R=R_S\) and \( \Omega=\Omega_S\) constants is equivalent to assuming a sufficiently large supply of nutrients, calcium and carbonate by the sea currents, in such a way that their depletion downstream is negligible. We will see later that the model captures some fundamental spatiotemporal dynamics of reef formation under this approximation. A more complete model including the dynamics of \( R\) and \( \Omega\) will be considered elsewhere.

Eqs. (\ref{eq:O-dtP}) and (\ref{eq:O-dtA}) encompass various biological, ecological, geological, and physical processes, whose intensity is regulated by different parameters, all positive. 
In the equation for the temporal evolution of living corals, Eq.~(\ref{eq:O-dtP}), we have included several terms accounting for: growth, $g_LRP$, nonlinear positive feedbacks in the coral, $fRP^2$ \cite{vandeLeemput2016}, and growth enhancement correlated with coral height due to light availability, $\ell ARP$ \cite{Izumi2023}. These three processes are mediated by the availability of resources $R$. We also account for adverse impacts on coral growth: basal coral mortality, $-mP$, saturation due to competition for space, $-sP^3$ \cite{Spencer2011}, and coral drying effects with reef height, $-dA^2P$ \cite{Gaffey1991}. Clonal reproduction leading to the colonization of space is represented by the nonlinear gradient term $g_C R|\bar{\nabla} P|^2$ and  the diffusion term $g_D\nabla^2 P$ \cite{PhysRevResearch.2.023402}.

For the temporal evolution of accumulated aragonite, Eq.~(\ref{eq:O-dtA}), we account for accretion generated by polyps (alive coral), described by $a(\Omega-1)P$ \cite{Mollica2018}, and erosion resulting from both biological and geological agents, $-eA$ \cite{Trudgill2011}. 

Rescaling time, space and the two variables of the system as 
    \begin{equation}
        \label{eq:resc-1}
        \begin{split}
            t' = g_L R_St, \quad \quad x' = \sqrt{\dfrac{g_L R_S}{g_D}}x,\\
            p = \dfrac{fP}{g_L} \quad \mathrm{and} \quad h = \dfrac{fR_S A}{a(\Omega_S-1)},
        \end{split}
    \end{equation}
we can reduce the number of relevant parameters in the model to six:    
    \begin{align}
        \begin{split}
            \partial_t p = &\Big[ (1-\alpha)+ p -\beta p^2 +\phi h -\gamma h^2  \Big]p \\
            &+\eta (\partial_x p)^2 + \partial_x^2 p,\label{eq:dtp}
        \end{split} \\
        \mathrlap{}
        \partial_t h = &p - \varepsilon h\ , \label{eq:dth}
    \end{align}    
where we have dropped the primes and we have defined
    \begin{equation}
        \label{eq:resc-2}
        \begin{split}
            \alpha = \dfrac{m}{g_LR_S}, \quad \beta = \dfrac{sg_L}{R_S f^2}, \quad \phi = \dfrac{a\ell (\Omega_S-1)}{g_L R_S f},\\
            \gamma = \dfrac{da^2(\Omega_S-1)^2}{g_LR_S^3f^2}, \quad \eta = \dfrac{g_C g_L R_S }{f g_D} \quad \mathrm{and} \quad \varepsilon = \dfrac{e}{g_L R_S}.
        \end{split}
    \end{equation}
Note that in the rescaled version of (\ref{eq:O-dtA}), Eq.~(\ref{eq:dth}), $A$ is named $h$, indicating our understanding of variable $A$ as an indirect measurement of the height the reef has reached from the ocean floor. 
        
All parameters in (\ref{eq:dtp}) and (\ref{eq:dth}) are positive. In addition, throughout all the work we consider $\Omega_S>1$, which implies that the medium is not sufficiently acidic to dissolve aragonite (a desirable scenario for coral reef formation which is threatened by the global change). As a result, all solutions of these equations are also positive, as long as the initial conditions are also positive. This is trivially proved for homogeneous solutions, as given $\partial_t p=0$ for $p=0$ and $\partial_t h \geq 0$ for $h=0$. In the following, we restrict, then, our analysis to positive values of $p$ and $h$, since negative values of these magnitudes have no ecological meaning. Throughout this paper we take $\beta=1/4$, $\phi=1$, $\gamma=1/4$, and consider $\alpha$, the rescaled mortality rate, and $\varepsilon$, the rescaled aragonite erosion rate, as our main control parameters. As it will be discussed later, this specific selection of less critical parameters does not alter the primary findings of this study, as the results presented remain robust across broad ranges of these parameters. The values chosen here simply facilitate a manageable and practical range for analyzing the other quantities under study. For simplicity, we also take $\eta=0$. We expect that this simplification does not significantly change the results of this paper, since it has been reported that the clonal term with which this parameter is associated induces only slight changes in the velocity of propagation of the spatially extended structures \cite{PhysRevResearch.2.023402}.

\section{\label{sec:Homogeneous Steady States} Homogeneous Steady States \protect}

In order to gain some insight into the dynamics of the set of Eqs. (\ref{eq:dtp}-\ref{eq:dth}), we analyze first its homogeneous steady states (i.e., fixed points) and study their stability. 
Bare soil, $\mathcal{S}_0 = (0,0)$, is always a solution of the system and populated states are given by 
\begin{equation}
        \label{eq:fix-p-3}
       \mathcal{S}_\pm =
       \begin{pmatrix}
            p_\pm \\ h_\pm
        \end{pmatrix}
        = h_\pm
       \begin{pmatrix}
            \varepsilon \\ 1
        \end{pmatrix},
\end{equation}
where
    \begin{equation}
        \label{eq:fix-p-2}
        h_\pm = \dfrac{(\phi+\varepsilon) \pm \sqrt{ (\phi+\varepsilon)^2 +4(1-\alpha)(\gamma+\beta\varepsilon^2)}}{2(\gamma+\beta\varepsilon^2)}.
    \end{equation}
See Appendix~\ref{app:Multiscale analysis} for more details on the computation of these solutions and their corresponding eigenvalues, from which the forthcoming linear stability analysis is derived.\\

\begin{figure*}[ht]
        \includegraphics[width=0.75\textwidth]{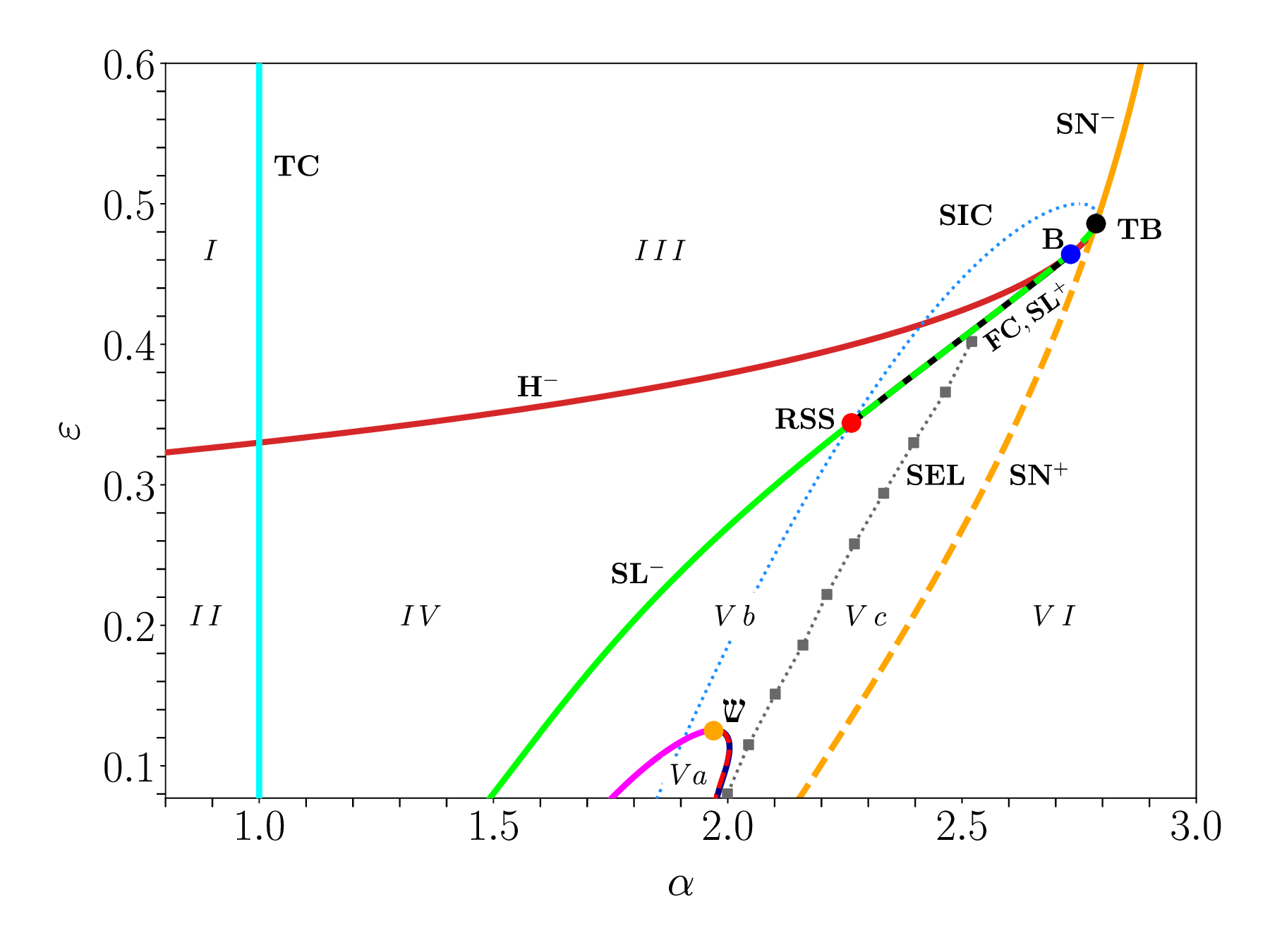}
        \caption{\label{fig:Bifurcations} Phase diagram of Eqs.~(\ref{eq:dtp}-\ref{eq:dth}) as a function of the mortality, $\alpha$, and erosion, $\varepsilon$, rates. Bifurcation lines divide the phase diagram in 8 regions with different dynamical regimes. To the left of the Transcritical bifurcation line ($\mathrm{TC}$, cyan), bare soil is unstable and homogeneous populations of live coral prevail, stationary in Region I, and oscillating in Region II. Regions I and II are separate by the Hopf bifurcation line  ($\mathrm{H^-}$, red). To the right of $\mathrm{TC}$ bare soil coexists stably with other solutions up to the Saddle Node bifurcation line ($\mathrm{SN}$, orange).
        Solid lines, in correspondence with labels with a $-$ superindex, indicate that the bifurcation creates one stable solution. Dashed lines (labels with $+$ superindex) indicate that the bifurcation creates only unstable solutions. In Region III there is bistability between the homogeneous populated solution and bare soil. In Region IV there is coexistence between an oscillatory populated solution and bare soil. In Region V, between the Saddle Loop line ($\mathrm{SL}$, green) and $\mathrm{SN}$ the system presents excitable behavior. The excitable region is divided in three subregions: In Region Va, enclosed by the violet and dark blue lines, explained further in Fig.~\ref{fig:Bifurcations-Zoom-Pulses}, one observes stable traveling pulses; in Region Vb, between  $\mathrm{SL}$ and the Spatial Excitability Limit ($\mathrm{SEL}$), persistent disordered structures are observed; and in Region Vc, between $\mathrm{SEL}$ and $\mathrm{SN}$ transient excitations decay to bare soil. The Saddle Index Change ($\mathrm{SIC}$, blue) is shown as a dotted thin line indicating it is not an actual bifurcation. In Region VI, to the right of $\mathrm{SN}$ the only possible state is bare soil. Bifurcation codimension-2 points are indicated with bold circles: Resonant Side Switching ($\mathrm{RSS}$, red), Bautin ($\mathrm{B}$, blue), Takens-Bogdanov ($\mathrm{TB}$, black) and $\shin$-point (orange). Some lines, such as the Fold of Cycles ($\mathrm{FC}$, black) cannot be properly labelled in this figure due to the scale. More details on the bifurcation structure around the $\mathrm{TB}$ point are presented in the zoom shown in Fig.~\ref{fig:Bifurcations-Zoom}. A zoom around the $\shin$-point is also shown in Fig.~\ref{fig:Bifurcations-Zoom-Pulses}. Other parameters: $\phi=1$ and $\beta=\gamma=1/4$.
        }
    \end{figure*}

The bare state $\mathcal{S}_0$ is stable for $\alpha > \alpha_T=1$ (and unstable for  $\alpha < \alpha_T$), where it undergoes a transcritical bifurcation (TC) with $\mathcal{S}_-$, as $p_-=h_-=0$ at the bifurcation. For $\alpha < 1$, $p_-$ is negative and it has no ecological meaning. Thus, we simply say that $p_-$ does not exist for $\alpha<1$. Fig.~\ref{fig:Bifurcations} shows the phase diagram of the system in the parameters space ($\alpha$, $\varepsilon$), where the bifurcation line corresponding to the TC is drawn in cyan color. This line separates regions I and II where bare soil is an unstable solution from the rest of the regions where it is a stable solution.

The populated solutions exist only for 
\begin{equation}
        \label{eq:alpha-SN}
        \alpha < \alpha_{SN} = 1 +\dfrac{(\phi+\varepsilon)^2}{4(\gamma+\beta\varepsilon^2)},
\end{equation}
where $\alpha_{SN}$ corresponds to the saddle-node bifurcation (SN) where 
$p_+=p_-$. For values of $\alpha$ larger than $\alpha_{SN}$ the only homogeneous solution of the system is bare soil. Bifurcation line corresponding to the SN is drawn in orange color in Fig.~\ref{fig:Bifurcations}. This line separates region VI where the only posible and stable solution is bare soil from the rest of the parameters space.

$\mathcal{S}_-$ exist for $\alpha\in(\alpha_T,\alpha_{SN})$ and is always a saddle point, whose stable manifold acts as a separatrix in phase space determining which initial conditions will evolve towards the highly populated solution $\mathcal{S}_+$ or towards bare soil. $\mathcal{S}_-$ undergoes a saddle-index change (SIC) at $\alpha=\alpha_{SIC}$ for which  
 \begin{equation}
        \label{eq:SIC-Condition}
        (1 -2\beta p_- )p_- -\varepsilon = 0.
\end{equation}
The SIC is indicated in Fig.~\ref{fig:Bifurcations} as a dotted blue line.
At the SIC, the ratio between the eigenvalues of $p_-$ changes, in absolute value, from being smaller to larger than 1 or vice versa, i.e. if at one side of the SIC line the negative eigenvalue is larger in absolute value than the positive eigenvalue, at the other side is the other way around. The SIC is not an actual bifurcation, as the stability of $\mathcal{S}_-$ does not change, but this line will be proven relevant for the dynamics later.

The highly population solution, $\mathcal{S}_+$, is stable in 
Regions I and III of the parameter space ($\alpha$, $\varepsilon$) (see Fig.~\ref{fig:Bifurcations}) and undergoes a Hopf bifurcation at $\alpha=\alpha_H$ for which 
    \begin{equation}
        \label{eq:HopfCondition}
        (1 -2\beta p_+ )p_+ -\varepsilon = 0.
    \end{equation}
The Hopf bifurcation line is drawn in red in Fig.~\ref{fig:Bifurcations}. $\mathcal{S}_+$ is unstable below this line, from which a limit cycle emerges, being this cycle stable only in regions II and IV of the parameters space. Close to the Hopf the frequency of the cycle is given by the imaginary part of the eigenvalues at the Hopf bifurcation (see Appendix \ref{app:Multiscale analysis}).
Explicit expressions for both $\alpha_{SIC}$ and $\alpha_H$ (the values at which Eq.~(\ref{eq:SIC-Condition}) and Eq.~(\ref{eq:HopfCondition}) are fulfilled, respectively) are also given in Appendix \ref{app:Explicit}.

\section{\label{sec:Global Bifurcations} Global bifurcations, higher codimension points, and excitability\protect}

\begin{figure}[h!]
    \includegraphics[width=\columnwidth]{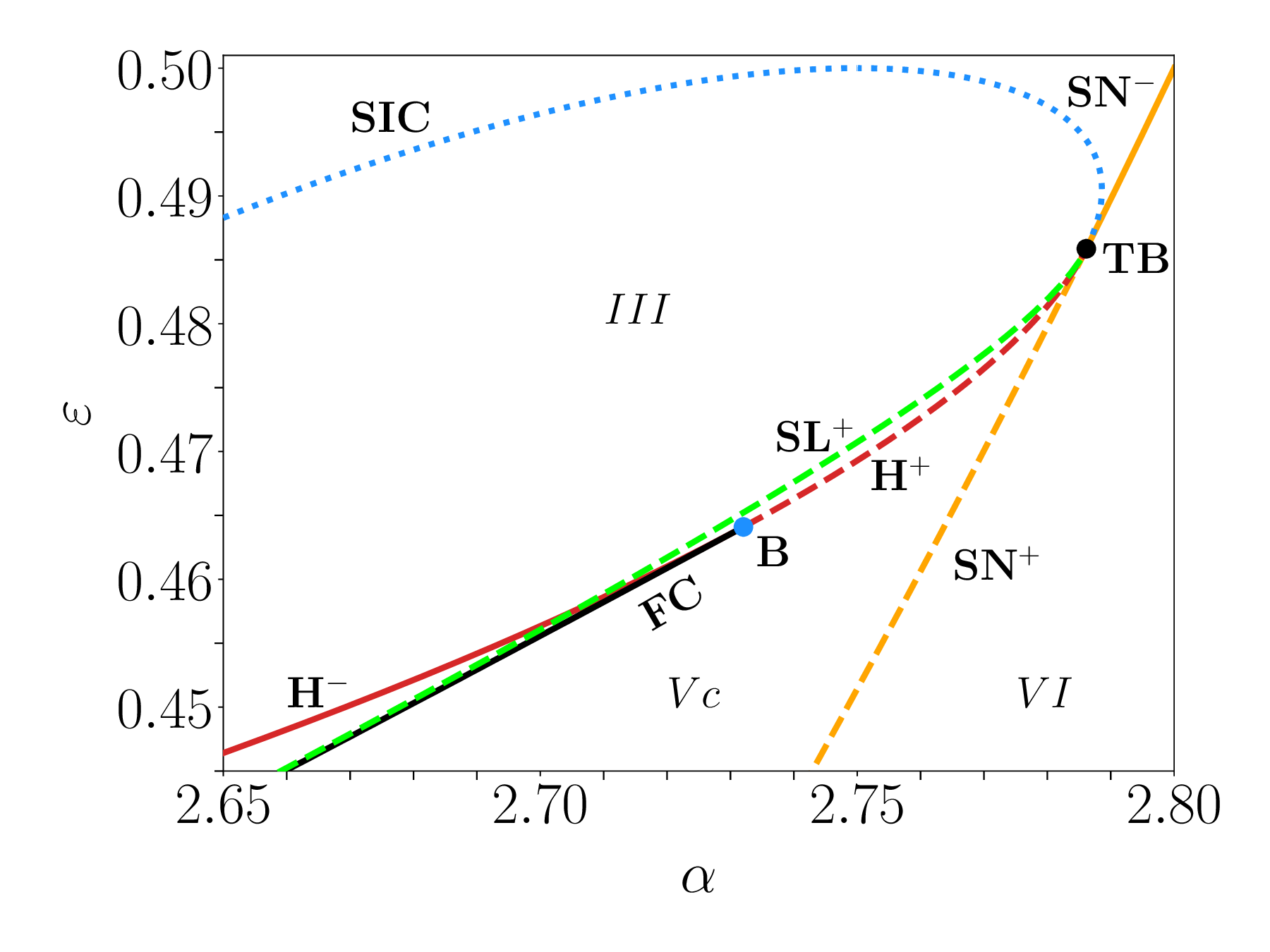}
    \caption{\label{fig:Bifurcations-Zoom} Zoom of Fig.~\ref{fig:Bifurcations} near the TB point. Saddle Index Change ($\mathrm{SIC}$, blue) is depicted as a dotted thin line indicating it is not an actual bifurcation. Bifurcation lines shown: unstable Saddle Node ($\mathrm{SN^+}$, orange dashed), stable Saddle Node ($\mathrm{SN^-}$, orange solid), Hopf of unstable cycle ($\mathrm{H^+}$, red dashed), Hopf of stable cycle ($\mathrm{H^-}$, red solid), Saddle Loop of unstable cycle ($\mathrm{SL^+}$, green dashed), Fold of Cycles  ($\mathrm{FC}$, black). Bifurcation co-dimension two points: Bautin ($\mathrm{B}$, blue) and Takens-Bogdanov ($\mathrm{TB}$, black). }
    \end{figure}

Hopf, Saddle Index Change, and Saddle-Node lines meet tangentially at a codimension-2 Takens-Bogdanov (TB) bifurcation. As the Hopf bifurcation destabilises the highly populated solution $\mathcal{S}_+$, the TB bifurcation divides the saddle node line, given by Eq.~(\ref{eq:alpha-SN}), in two parts: $\mathrm{SN}^+$ from which $\mathcal{S}_+$ emerges unstable, and $\mathrm{SN}^-$ from which  emerges  stable. For the set of parameters used in this paper, the cycle associated to the Hopf bifurcation near the TB bifurcation is unstable. Thus we label the Hopf line as $\mathrm{H}^{+}$ and indicate it in Fig.~\ref{fig:Bifurcations-Zoom} as a dashed red line. This cycle is destroyed at a Saddle- Loop (SL) or Homoclinic bifurcation,  where it collides with $\mathcal{S}_-$ becoming a homoclinic orbit. Since the cycle is unstable, we name this bifurcation $\mathrm{SL}^+$ and indicate it as a dashed green line in Figs.~\ref{fig:Bifurcations} and \ref{fig:Bifurcations-Zoom}. As expected, the Saddle-Loop bifurcation also emerges tangentially from the TB bifurcation  \cite{guckenheimer1983nonlinear}.

Concerning the critical values of $\alpha$ and $\varepsilon$ at the TB point, they are obtained by equating Eqs.~(\ref{eq:alpha-SN}) and (\ref{eq:alfa_SIC-Hopf}), solving for $\varepsilon$ and then inserting the obtained value in either Eq.~(\ref{eq:alpha-SN}) or (\ref{eq:alfa_SIC-Hopf}) in order to get the corresponding value of $\alpha$. Solving numerically for the set of parameters used in this paper, we obtain: $\alpha_{TB}=2.768151(4)$ and $\varepsilon_{TB}=0.485868(3)$.
    
The stability of the cycle at the Hopf bifurcation changes in the codimension-2 Bautin point (B), turning to a stable cycle. Therefore, the Hopf bifurcation line beyond the Bautin point is now labelled $\mathrm{H}^{-}$. Associated with the Bautin point, there is a fold (or saddle-node) of cycles line (FC) indicated with a solid black line in Figs.~\ref{fig:Bifurcations} and \ref{fig:Bifurcations-Zoom}, where the stable cycle that emerges from $\mathrm{H}^{-}$ coalesces with the unstable cycle emerging from $\mathrm{SL}^{+}$ and both are destroyed. The Bautin point is located at $\alpha_B= 2.732050(8)$ and $\varepsilon_B = 0.464101(6)$. Check Appendix \ref{app:Multiscale analysis} for further details on this point.

Finally, the FC and the $\mathrm{SL}^+$ lines enter tangentially the codimension-2 Resonant Side Switching (RSS) point \cite{PhysRevA.77.033841,Chow,Champneys-Kuznetsov}, and they coalesce to a single Saddle Loop bifurcation line of the stable cycle ($\mathrm{SL}^-$), indicated as a solid green line in Fig.~\ref{fig:Bifurcations}. RSS occurs when the saddle point of the homoclinic trajectory associated to the SL undergoes a saddle index change. That is precisely the reason why we determined the expression for the SIC line (\ref{eq:SIC-Condition}) in the previous section. For the set of parameters used in this paper, this point is located at $\alpha_\mathrm{RSS}=2.264(1)$ and $\varepsilon_\mathrm{RSS}=0.344(0)$.

\begin{figure}[h!]
    \includegraphics[width=\columnwidth]{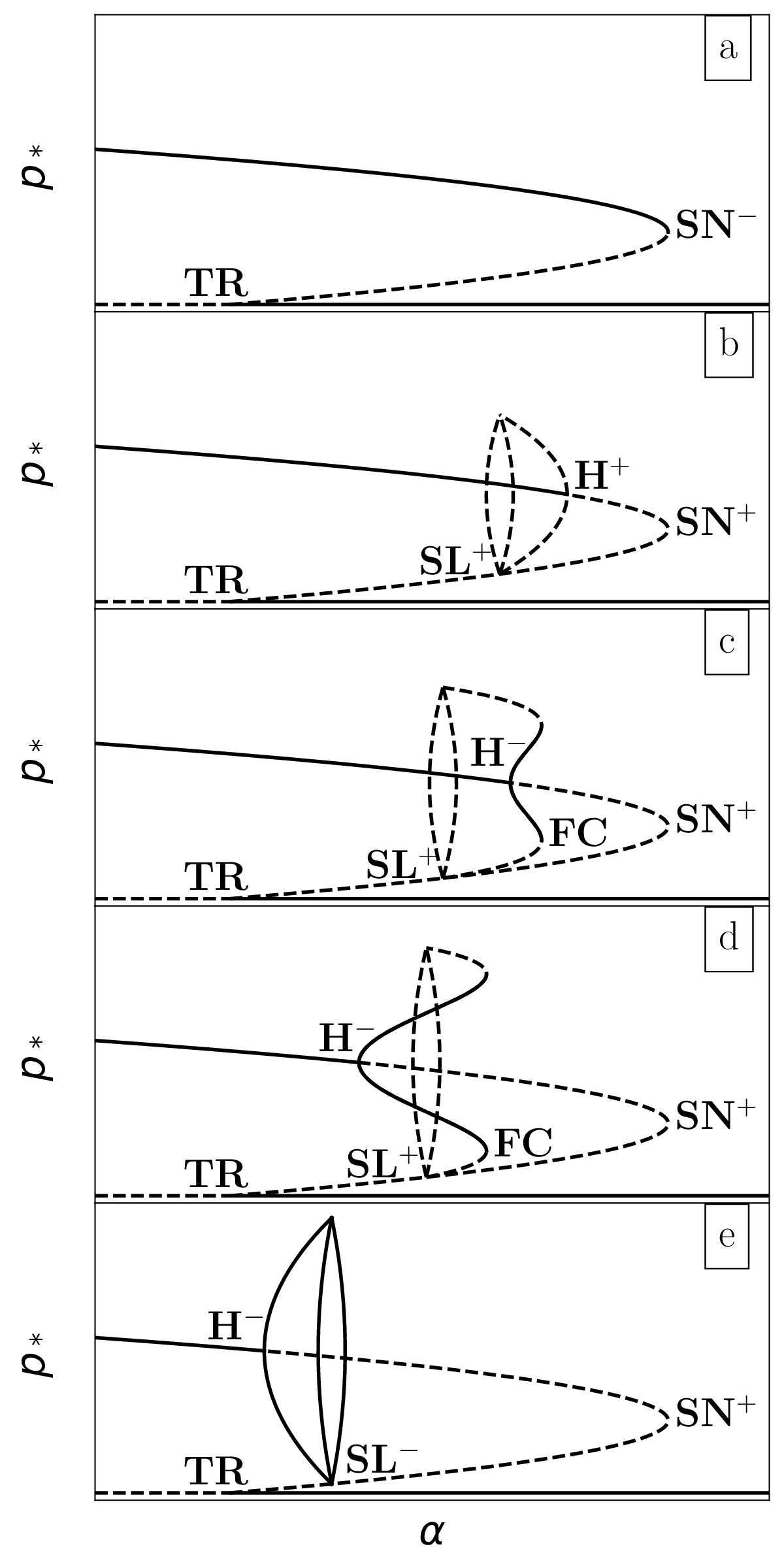}
    \caption{\label{fig:P_stationary} Behaviour of stationary homogeneous solutions $p_*$ of the system for a specific choice of some parameters ($\phi=1$, $\beta=\gamma=1/4$) and different values of $\varepsilon$ as a function of $\alpha$. Both vertical and horizontal axes are in arbitrary units. Plots are not at real scale for the sake of clarity. Solid (resp. dashed) lines represent stable (resp. unstable) fix points or cycles. Bifurcation points shown: Transcritical ($\mathrm{TC}$), stable and unstable Saddle Node ($\mathrm{SN^-}$ and $\mathrm{SN^+}$, respectively), Hopf of stable and unstable cycle ($\mathrm{H^-}$ and $\mathrm{H^+}$, respectively), Fold of Cycles ($\mathrm{FC}$) and Saddle Loop of stable and unstable cycle ($\mathrm{SL^-}$ and $\mathrm{SL^+}$, respectively). The correspondent values of $\varepsilon$ for the different panels are a) $\varepsilon=0.490$, b) $\varepsilon=0.470$, c) $\varepsilon=0.462$, d) $\varepsilon=0.450$ and e) $\varepsilon=0.340$.}
    \end{figure}

To illustrate the complex bifurcation scenario described above we plot in Fig.~\ref{fig:P_stationary} several sketches of characteristic bifurcation diagrams of $p^*$ as a function of $\alpha$ corresponding to horizontal cuts of Figs.~\ref{fig:Bifurcations} and \ref{fig:Bifurcations-Zoom} for decreasing values of $\varepsilon$. 

We note that for values of $\alpha$ beyond the Saddle-Loop bifurcation in Fig.~\ref{fig:P_stationary}e (Regions Va, Vb, and Vc in Figs. \ref{fig:Bifurcations}, \ref{fig:Bifurcations-Zoom}, and \ref{fig:Bifurcations-Zoom-Pulses}) the system displays Type-I temporal excitability \cite{Rinzel1989,Izhikevich2000,Izhikevich2007}. In this regime the only stable steady state is bare soil, and low polyp densities will decay exponentially to zero. Densities above $p_-$, however, will grow following the remainings of the destroyed cycle in a large excursion in phase space, to eventually go back to zero. During the growth stage polyps accumulate aragonite but once $h$ reaches the sea level polyps start dying due to drying. As a result, aragonite accumulation can not replenish the eroded rock, and both $p$ and $h$ decay to zero. This temporal behavior of homogeneous solutions has implications for the spatiotemporal dynamics of coral reefs as discussed in the next Section.

\section{\label{sec:TP} Dynamics of traveling pulses: reef formation\protect}

\begin{figure}[h!]
    \includegraphics[width=\columnwidth]{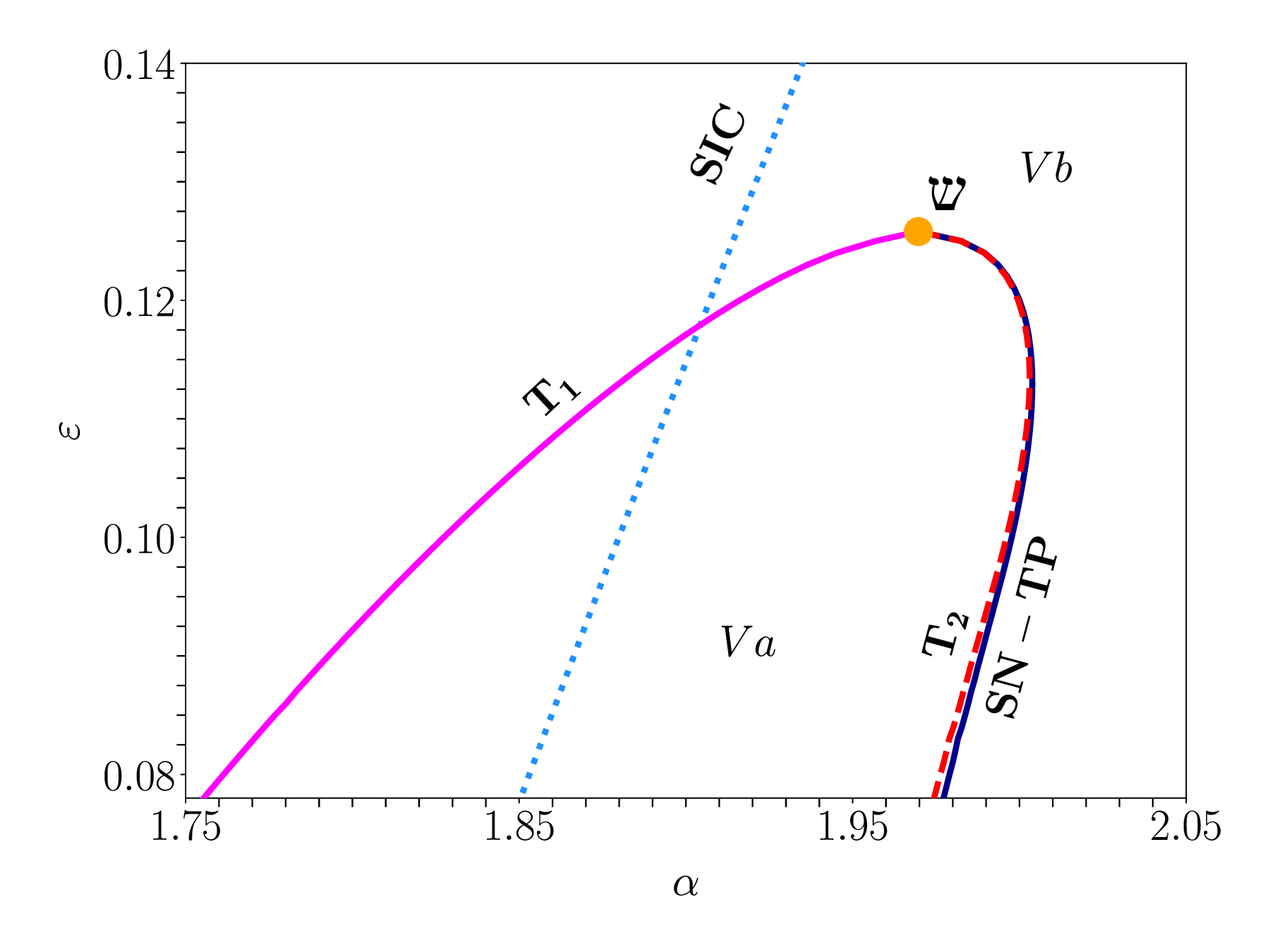}
    \caption{\label{fig:Bifurcations-Zoom-Pulses} Zoom of Fig.~\ref{fig:Bifurcations} near the $\shin$ point. Saddle Index Change ($\mathrm{SIC}$, blue) is depicted as a dotted thin line indicating it is not an actual bifurcation. Bifurcation lines shown: line of T-points on the stable pulse ($\mathrm{T_1}$, magenta), line of T-points on the unstable pulse ($\mathrm{T_2}$, red dashed), and Saddle Node of traveling Pulses (SN-TP, dark blue). Bifurcation co-dimension two point: $\shin$-point ($\shin$, orange). Traveling pulses are stable in Region Va. }
    \end{figure}

The dynamics of homogeneous solutions alone is insufficient to account for the observed shapes and evolution of coral reefs. While the bifurcations discussed in the previous section define and structure the existence and stability of stationary and oscillatory homogeneous solutions within parameter space, further analysis is required to describe spatially extended, non-homogeneous dynamics. Notably, due to the terms involving spatial derivatives, beyond a Saddle-Loop bifurcation temporal excitable excursions can give rise to spatially extended structures that propagate \cite{Arinyo-i-Prats2021,Moreno-Spiegelberg2022}. These solutions emerge within the excitable region of parameter space  and, unlike their homogeneous counterparts, do not decay to zero over time in Regions Va and Vb of Figs. \ref{fig:Bifurcations}, \ref{fig:Bifurcations-Zoom}, and \ref{fig:Bifurcations-Zoom-Pulses}.  
Additionally, the existence of such structures has been suggested as a potential mechanism through which ecosystems can enhance their resilience to stressors and high mortality rates \cite{Moreno-Spiegelberg2024}. In Region Vc, excitable excursions cause only transient boosts of coral that eventually decay back to the rest state.

\begin{figure}[h!]
    \includegraphics[width=\columnwidth]{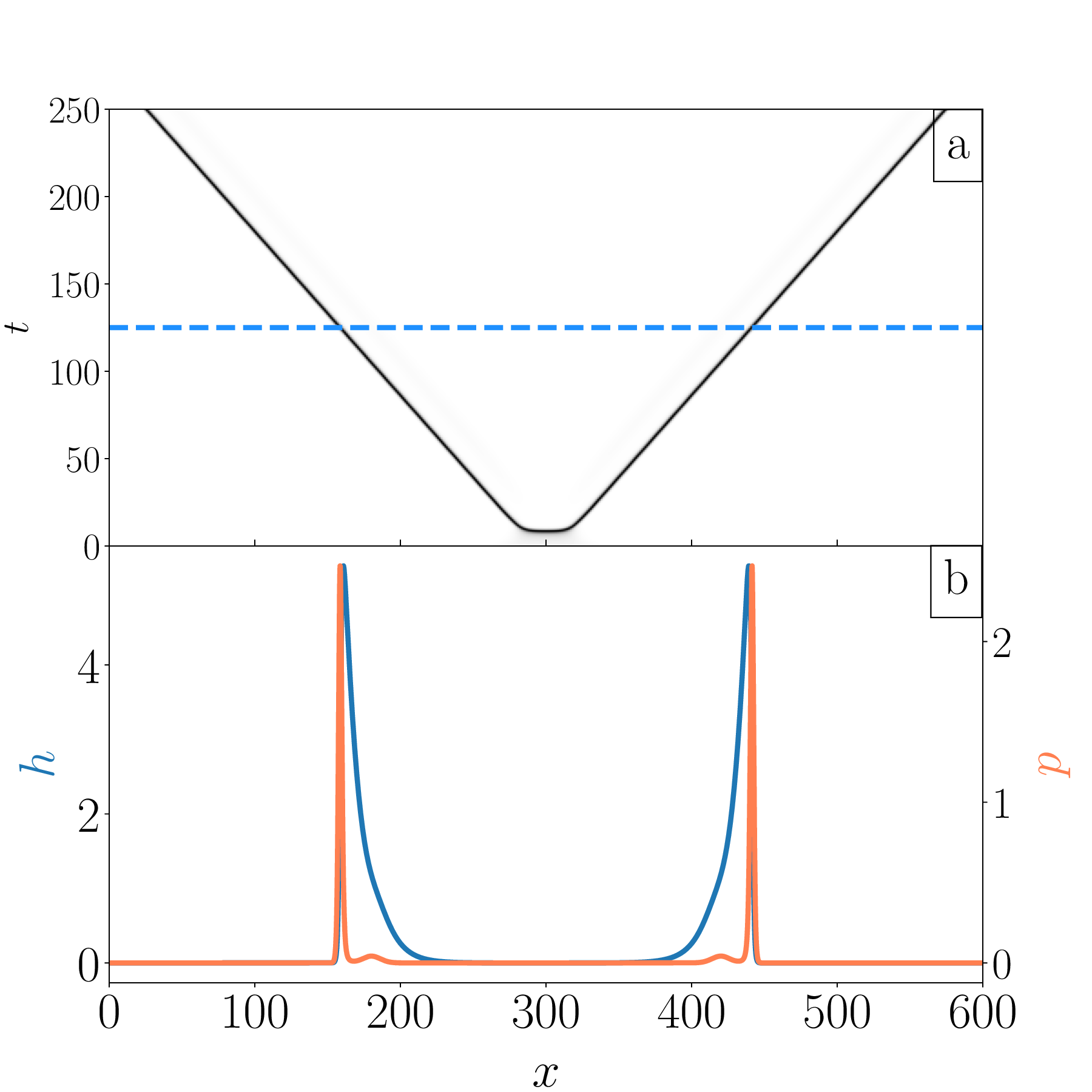}
    \caption{\label{fig:PulseStable} Stable pulses traveling in space for $\alpha=1.85$ and $\varepsilon=0.1$. The two counter propagating pulses were formed starting from an initial condition given by Eq.~(\ref{eq:Supergaussian}) with $L=600$ and $\sigma=15.63$. a) Space-time representation of the TP dynamics (darker colors denote higher values of $p$ and white denotes $p=0$). The dashed line indicates the time corresponding to the cross-section shown in panel b). b) Shape of the pulse at time $t=125$. Field $p$ is depicted in orange and field $h$ is depicted in blue.}
    \end{figure}

Our study primarily focuses on one type of spatially non-homogeneous solutions -- traveling pulses (TPs) -- as they capture the fundamental dynamics of coral reef systems. These solutions are stable in Region Va. TPs retain their spatial profile while propagating through space at a constant velocity \( c \). We identify stable TPs (see Fig.~\ref{fig:PulseStable}) in the form of large structures of alive coral and aragonite that travel over bare soil. Coral grows on the upstream slope of the rock, expanding it in one direction by accumulating aragonite, while erosion destroys the rock on the back, leaving the shape of the traveling pulse unchanged. 

The two TPs in Fig.~\ref{fig:PulseStable}, moving in opposite directions, were generated from a super-Gaussian initial condition, Eq.~(\ref{eq:Supergaussian}), where the peak values of \( p \) and \( h \) slightly exceeded the threshold \( \mathcal{S}_- \). The dynamics unfold as follows: initially, live coral colonies grow and accumulate aragonite, allowing the reef to develop in place until it reaches the water surface. At that point, coral colonies begin to die at the center due to high temperatures and drying, but they continue to grow along the outer, submerged slopes of the hard rock, causing lateral reef expansion. Over time, the exposed central aragonite erodes, creating two distinct pulses that move in opposite directions.

These pulses lose their stability through two types of bifurcations: the Saddle Node of Traveling Pulses (SN-TP) and the T-point (T$_1$) (Fig. \ref{fig:Bifurcations-Zoom-Pulses}). In the SN-TP bifurcation the stable TP collides with an unstable TP, and there are no TP beyond this point. The T-point, on the other hand, is a distinct bifurcation, also known as a heteroclinic bifurcation. Near this point, TPs begin to form a large plateau that approaches the values of the low-population state $\mathcal{S}_-$. As the system nears the T-point, the width of this plateau expands until it diverges at the T-point, expanding over the entire spatial domain. Beyond the T-point, a new unstable TP exhibits a spatial profile resting entirely on the $\mathcal{S}_-$ fixed point. For a more detailed explanation of this bifurcation, see \cite{Knut-Alfsen_1985, Glendinning1986, Moreno-Spiegelberg2022, Or-Guil2001}.
   
We also identify the existence of a second T-point (T$_2$) for the unstable TPs that emerge from the SN-TP bifurcation. Similar to the behavior of the stable TP near T$_1$, the unstable TP near T$_2$ begins to develop an extended tail, which eventually takes over the entire spatial domain at the bifurcation point. The new TPs emerging from T$_2$ rest also on the saddle \( \mathcal{S}_- \) and are unstable, but exhibit a smaller amplitude compared to the unstable TP originated from T$_1$. By employing continuation methods to track the TPs in Eqs.~(\ref{eq:dtp}-\ref{eq:dth}) (see Appendix \ref{app:Methods} for further details), we have located these bifurcations in parameter space, marked by the SN-TP, T$_1$ and T$_2$ lines in Figs.~\ref{fig:Bifurcations} and \ref{fig:Bifurcations-Zoom-Pulses}.

Lines  SN-TP, T$_1$ and T$_2$ meet in a high codimension point for which we are not aware of previous reports in the literature. We label this point with the hebrew letter $\shin$ (shin), based on the shape of the bifurcation lines around it. This point is located at $\alpha_\shin$=1.969(7) and $\varepsilon_\shin$=0.125(8). 

TPs are stable in the parameter region enclosed by the $\mathrm{T_1}$ and SN-TP lines (Region Va). Other spatio-temporal structures can also co-exist in this region, for instance, traveling waves formed by sufficiently well-spaced TPs.

Regarding the behavior of TPs outside the stability region, it is important to note that when the value of \( \alpha \) decreases after crossing the \( \mathrm{T}_1 \) point, the unstable pulse velocity increases until it diverges at the \( \mathrm{SL}^- \) bifurcation. This divergence causes the TPs to travel so rapidly that they effectively transform into homogeneous solutions, following the dynamics of the homoclinic bifurcation at \( \mathrm{SL}^- \) \cite{Moreno-Spiegelberg2022}.

Conversely, if we leave the stability region by following the branch of unstable pulses originating from the \( \mathrm{T}_2 \) point (increasing \( \alpha \)), we observe that both the pulse’s amplitude and velocity decrease until reaching zero at the saddle node of unstable homogeneous solutions (\( \mathrm{SN}^+ \)). Beyond this bifurcation, TPs no longer exist.

\begin{figure}[h!]
        \includegraphics[width=\columnwidth]{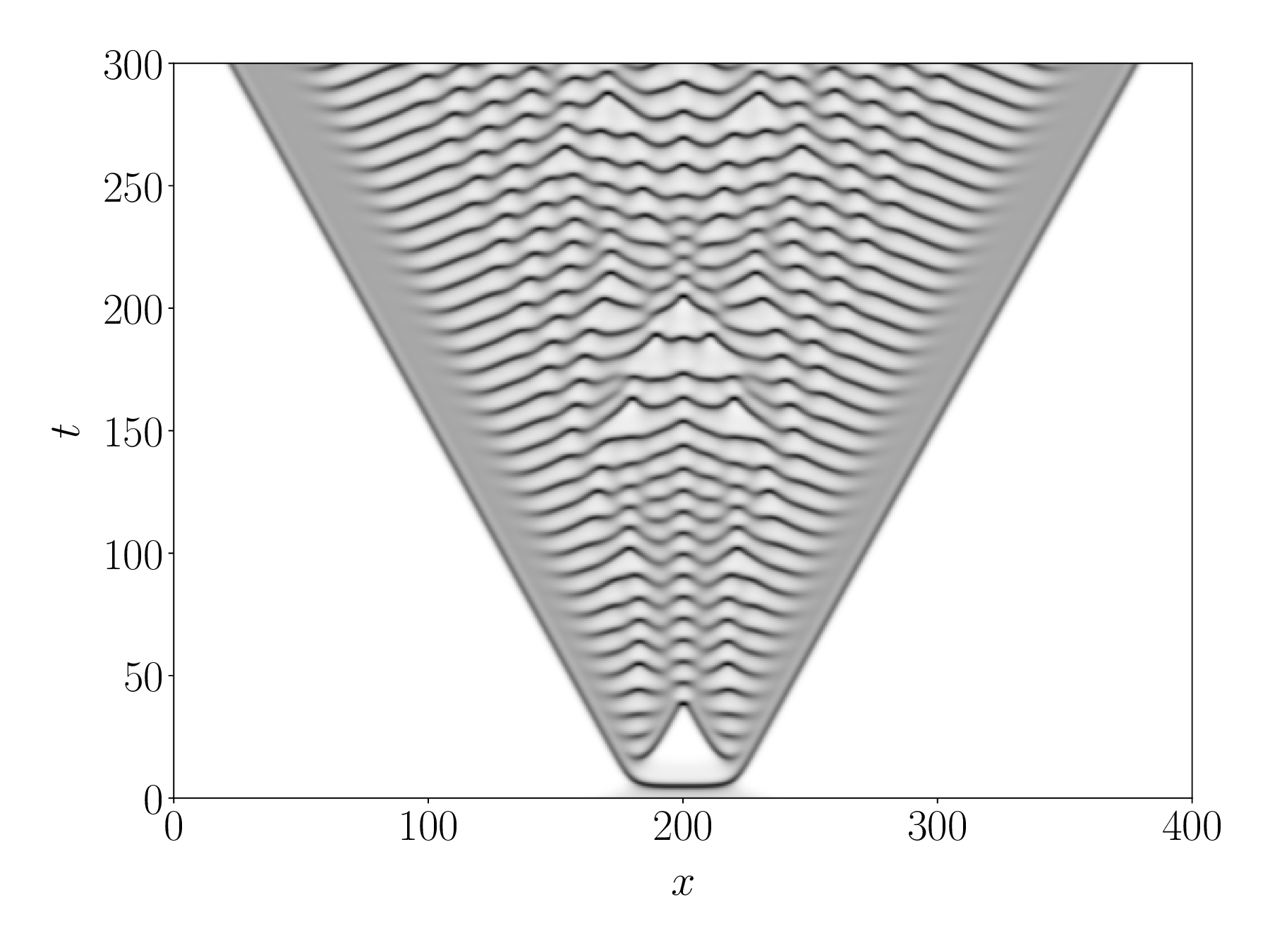}
        \caption{\label{fig:Turbulence} Self-sustained spatially extended solution traveling in space for $\alpha=2.2$ and $\varepsilon=0.3$. Initial condition given by Eq.~\ref{eq:Supergaussian} with $L=600$ and $\sigma=15.63$. We have not determined whether this situation corresponds to a chaotic regime or to a superposition of traveling waves.}
    \end{figure}

The instability of TPs does not imply that any spatially heterogeneous perturbation of the bare state \( \mathcal{S}_0 \) will always decay over time.  As mentioned earlier, there may be stable traveling waves that extend their stability beyond the limits of the TPs stable region. Additionally, more complex self-sustained dynamics, such as spatiotemporal chaos, are present across a wide region of the parameter space. An example of such solutions is shown in Fig.~\ref{fig:Turbulence}. While there are clear similarities between the dynamics shown in Fig.~\ref{fig:PulseStable} and the one in Fig.~\ref{fig:Turbulence}, there are also distinct differences: the former maintains a regular spatial profile over time, while the latter, although exhibiting an outer boundary reminiscent of a TP, shows intricate spatiotemporal dynamics within its inner region. These kinds of solutions can be of particular relevance in the context of coral reef dynamics, as they may describe atolls with intricate coral structures inside.

We have investigated the time evolution of initial super-Gaussian profiles across various values of \( \alpha \) and \( \varepsilon \) to estimate the region of parameter space where the system exhibits spatial structures characterized by self-sustained dynamics. By observing whether the initial condition decays to the bare state \( \mathcal{S}_0 \) or travels or expands over time, we determined the Spatial Excitability Limit (SEL) — the boundary in parameter space separating the region that displays self-sustained dynamics (Region Vb) from the region that do not (Region Vc). The shape of this boundary is somewhat irregular, likely due to the inherent arbitrariness in the choice of the initial condition and the time frame for checking whether \( p \) and \( h \) remain non-zero. For this reason, we just present several points of the SEL in Fig.~\ref{fig:Bifurcations}, connected by a dotted line for visual guidance. Self-sustained spatio-temporally disordered dynamics is observed to the left of the SEL line in Fig.~\ref{fig:Bifurcations}, all the way to the $\mathrm{SL}$ line.

   \section{\label{sec:Discussion}Discussion\protect}

The comprehensive analysis of the bifurcations in our model lays the groundwork for its applications to coral reef formation studies. This model successfully replicates the formation and development of reefs without requiring external factors beyond the interactions between the living coral, its environment (specifically resource uptake, space occupation, light availability, and drying risk), and its intrinsic processes, such as Allee effects and aragonite accretion with the resulting lateral and vertical expansion. 
   
Specifically, we can relate the two excitable pulses travelling in opposite directions to closed atolls in two spatial dimensions which are empty inside  (Fig. \ref{fig:PulseStable}). As a matter of fact, TPs are stable in the lower part of Fig.~\ref{fig:Bifurcations}—for low $\varepsilon$ values. This is the region of parameter space where such structures would be expected to form, as high erosion values would lead to the degradation of calcareous structures and the eventual disappearance of the atoll as a spatial feature. Instead, this would result in a more or less homogeneous and dispersed coral distribution, incapable of building substantial aragonite structures. 

We can also interpret the self-sustained spatial structures with entangled internal dynamics (Fig.~\ref{fig:Turbulence}) as closed atolls showing labyrinthine coral structures in their inner lagoons. In our model, such structures form for slightly larger erosion rates or lower mortalities. Fringing reefs,  i.e. a single coral barrier with no back reef, correspond in our description to a single travelling pulse. This can be formed by suitable initial conditions, for instance, if the initial coral colonies are very close to the shore and a back reef can not form. 
    
One-peaked solutions can also form in environments with heterogeneity in the resources caused by advection. If the advective current flows from right to left, it could happen that the first coral peak (the one traveling to the right) depletes so many resources that the effective local value of $\alpha$ in the place where the second peak (the one traveling to the left) should be located is pushed beyond the spatial excitability limit (SEL) and it dies. In simpler words, if the resource supply is too weak, depletion caused by the presence of alive coral in the front of the reef (as seen following the movement of ocean currents) makes the rear part of the reef uninhabitable for corals.

Excitable pulses have also been studied in populations of other sessile species \cite{Marasco2014,Fernandez-Oto2019,Iuorio2021,Ruiz-Reynes2023,Karst2016,Davidson1997}. Generally, creating excitable structures requires mechanisms involving positive and negative feedback mediated by an inhibitor \cite{Moreno-Spiegelberg2024}. In our model, negative feedback arises due to the proximity of the upper (living coral) part of the reef to the sea level due to aragonite accretion and the consequent reef vertical expansion. Nevertheless, this dynamics is nonlinear, because in the regions where the vertical growth of the reefs from the sea floor has been moderate --intermediate heights-- the living conditions for corals are enhaced by improving light availability. Thus, aragonite acts as a density-dependent activator/inhibitor, promoting re-excitation for moderate heights. We note that the dynamics of the living coral population also involve feedback mechanisms that shift from positive to negative due to Allee effects and space occupation. However, in the absence of aragonite, these mechanisms do not induce excitability but instead contribute to stabilizing both the bare and highly populated states.

We conducted preliminary simulations (not shown in this work) of the model, Eqs.~(\ref{eq:dtp}-\ref{eq:dth}), extended to two spatial dimensions, and found stable spirals in the spatial excitability region, similar to those reported in a related model in a different context \cite{Moreno-Spiegelberg2024}. This dynamics, which is absent in coral reefs, do not emerge when an advective supply of resources is included. Further analysis of an extension of the model in two spatial dimensions will be reported elsewhere. Also, a basic sensitivity analysis of the model to parameter changes suggests that the bifurcation structure described in this study is robust across a broad range of parameter values, provided they remain positive. While these findings are tentative, they indicate that the observed scenario is not a consequence of fine-tuning the model parameters.

Our findings reveal that a large parameter space region allows the model solutions to reproduce observed coral reef shapes without requiring  external factors, such as sea level fluctuations or pre-existing non-flat geological structures (though the model can be extended to include these factors, potentially enhancing its descriptive power). These insights challenge the classical Darwinian theory of coral reef formation \cite{DarwinsTheory}, which posits that reefs grow atop pre-existing geological structures, typically volcanic islands, with their shapes being a direct consequence of this foundation. We argue that the framework we propose encompasses a broader range of scenarios, extending beyond reefs built over volcanic remnants. 

It should also be noted that this model does not exhibit Turing instabilities that could offer alternative explanations to coral reef pattern formation. Homogeneous solutions are always stable against periodic perturbations (see Appendix \ref{app:Multiscale analysis} for further details). Nevertheless, we do not assert that Turing patterns are irrelevant to the description of coral reefs. Indeed, they have previously been utilized to describe periodic reefs \cite{Mistr-Bercovici}.
Although this pioneering work is highly valuable for the mathematical study of reef patterns, it does not encompass the range of scenarios our model illustrates. Therefore, we argue that traveling pulses and self-sustained spatiotemporal dynamics, along with accurate simulations of the resources supplied by water currents, offer an appropriate framework for the mathematical representation and explanation of a broad range of coral reefs.

\section{\label{sec:Conclusions}Concluding remarks\protect}

In conclusion, our study highlights the crucial role of excitable dynamics and the bifurcation structure of traveling pulses in modeling coral reef formation. Without invoking external factors such as geological formations or sea-level changes,
and by focusing merely on the interactions between living coral and its environment, we demonstrate how the model reproduces the development of various reef structures, ranging from atolls to fringing reefs. These patterns emerge from intrinsic feedback mechanisms within the reef system, such as resource uptake, aragonite accretion, and the Allee effect.

We have shown that traveling pulses, representing different reef formations, and complex spatiotemporal dynamics can explain the most general of coral reef morphologies. These structures form naturally as solutions to the governing equations, emphasizing the importance of self-sustained dynamics in coral ecosystems. Notably, our findings suggest that the resilience and spatial organization of reefs may be enhanced by such excitable processes, offering new insights into their stability and adaptability.

While this model challenges aspects of classical Darwinian theory by showing that reefs can form without pre-existing geological structures, it complements the traditional framework by providing a broader, more flexible approach to reef formation. Future extensions of the model, including the effects of water currents and more detailed resource dynamics, could offer further refinement and applicability to real-world coral ecosystems.

Ultimately, this research underscores the potential of excitable systems and bifurcation theory to offer a robust mathematical explanation for coral reef formation, providing a foundation for future studies on the resilience and evolution of coral reefs in the face of environmental stressors.

    \nocite{*}

    \providecommand{\noopsort}[1]{}\providecommand{\singleletter}[1]{#1}%

        \onecolumngrid
    \appendix

    \section{\label{app:Multiscale analysis} Multiple scales analysis of the spatially homogeneous solutions of the model\protect}

    In this section, we derive systematically a series of equations for the evolution of perturbations around the stationary homogeneous solutions of Eqs.~(\ref{eq:dtp}-\ref{eq:dth}) using the method of multiple scales. We introduce the small parameter $\epsilon\ll 1$ (do not confuse it with the $\varepsilon$ parameter of the model) as a measure of the scale of the perturbations
    \begin{equation}
        \label{eq:S-exp}
        \mathcal{S} = \mathcal{S}_* + \epsilon \mathcal{S}_1 + \epsilon^2 \mathcal{S}_2 + \epsilon^3 \mathcal{S}_3 + ... ,
    \end{equation}
    where $\mathcal{S}_*$ is any of the stationary solutions $\mathcal{S}_0,\mathcal{S}_+$ and $\mathcal{S}_-$ discussed in the main text. The other terms $\mathcal{S}_1$, $\mathcal{S}_2$, $\mathcal{S}_3$, ... are the solutions at different orders of $\epsilon$. Written explicitly in its components, this equation reads
    \begin{equation}
        \label{eq:S-exp-1}
        \mathcal{S} = \begin{pmatrix} p_* \\ h_* \end{pmatrix}
        + \epsilon \begin{pmatrix} p_1(t) \\ h_1(t) \end{pmatrix}
        + \epsilon^2 \begin{pmatrix} p_2(t) \\ h_2(t) \end{pmatrix}
        + \epsilon^3 \begin{pmatrix} p_3(t) \\ h_3(t) \end{pmatrix} + ...
    \end{equation}
    
    We select $\alpha$ as our control parameter and expand it in powers of $\epsilon$
    \begin{equation}
        \label{eq:alpha-exp}
        \alpha = \alpha_0 + \epsilon \alpha_1 + \epsilon^2 \alpha_2 + \epsilon^3 \alpha_3 + ... 
    \end{equation}

    Moreover, to introduce a hierarchy of time scales modulated by parameter $\epsilon$ in order to differentiate between slow and fast dynamics in the system
    \begin{equation}
        \label{eq:dt-exp}
        \partial_t = \partial_{\tau_0} + \epsilon \partial_{\tau_1} + \epsilon^2 \partial_{\tau_2}  + \epsilon^3 \partial_{\tau_3} ... 
    \end{equation}

    Using (\ref{eq:S-exp-1}), (\ref{eq:alpha-exp}), and (\ref{eq:dt-exp}) in $\partial_t \mathcal{S}$ and Eqs.~(\ref{eq:dtp}) and (\ref{eq:dth}), we obtain an equation for each power of $\epsilon$. We shall study each of these orders separately up to $\epsilon^3$.\\
    
    At order $\epsilon^0$, we have
    \begin{equation}
        \label{eq:O-eps-0}
        \partial_{\tau_0} \begin{pmatrix} p_* \\ h_* \end{pmatrix} =
        \begin{pmatrix}
            \big[(1-\alpha_0) + p_* +\phi h_* - \beta p_*^2 - \gamma h_*^2\big]p_* \\
            p_* - \varepsilon h_* 
        \end{pmatrix},
    \end{equation}
    the definition of the stationary states. The left hand side of this equation is identically zero since the stationary solutions have null derivatives with respect to every time scale ($\tau_0,\tau_1,\tau_2,...$). The right hand side of this equation determines the relations that stationary states must fulfil, whose solutions are $\mathcal{S}_0$ and $\mathcal{S}_\pm$ given in Sec.~\ref{sec:Homogeneous Steady States}.\\
    The first one of those is 
    \begin{equation}
        \label{eq:fix-h}
        p_*=\varepsilon h_*,
    \end{equation}
    which immediately gives the other two: $p_0=h_0=0$ and
    \begin{equation}
        \label{eq:fix-h-2}
         (\gamma+\beta\varepsilon^2)h_\pm^2 - (\phi+\varepsilon) h_\pm - (1-\alpha)  = 0.
    \end{equation}
    
    It is important to note that every piece of information obtained when working at some order in $\epsilon$ will be used in the analysis at subsequent orders. For instance, that $\mathcal{S}_*$ has null derivatives with respect to every time scale and that the term inside square brackets in Eq.~(\ref{eq:O-eps-0}) is zero for the $\mathcal{S}_\pm$ stationary states will be implicitly used from now on in order to make the text clearer.\\

    At order $\epsilon^1$, we have
    \begin{equation}
        \label{eq:O-eps-1}
        \partial_{\tau_0} \begin{pmatrix} p_1 \\ h_1 \end{pmatrix} = J(p_*,h_*) \begin{pmatrix} p_1 \\ h_1 \end{pmatrix} + \begin{pmatrix} - \alpha_1 p_* \\ 0 \end{pmatrix},
    \end{equation}
    where $J(p,h)$ is the Jacobian of the system of equations (\ref{eq:dtp}-\ref{eq:dth})
    \begin{equation}
         \label{eq:Jac}
        J(p,h)=
        \begin{pmatrix}
            (1-\alpha)+ (2-3\beta p)p +(\phi -\gamma h) h & \big(\phi -2\gamma h  \big)p\\
            1 & -\varepsilon
        \end{pmatrix}.
    \end{equation}

    We can infer the linear stability of the stationary solutions from the eigenvalues of $J$. We first look at the simple case of the bare state $\mathcal{S}_0$. The Jacobian evaluated at this point is 
    \begin{equation}
        \label{eq:Jac-0}
        J(p_0,h_0)=
        \begin{pmatrix}
            1-\alpha_0 & 0\\
            1 & -\varepsilon
        \end{pmatrix},
    \end{equation}
    whose eigenvalues are trivially $\lambda_\alpha = 1-\alpha$ and $\lambda_\varepsilon = -\varepsilon$.\\
    Since $\varepsilon>0$ we have always that $\lambda_\varepsilon<0$. Hence, the stability of $\mathcal{S}_0$ is determined by the value of $\alpha$. If $\alpha>1$, both eigenvalues are negative and the fixed point is stable. If $\alpha<1$ then $\lambda_\alpha$ is positive and the fixed point is a saddle. We then conclude that the only bifurcation this solution undergoes is a transcritical one for $\alpha_{TC}=1$. We previously inferred this result in Sec.~\ref{sec:Homogeneous Steady States} where we studied the dependence of the sign of the solution $p_-$ dependence on $\alpha$.\\

    The Jacobian evaluated at the populated solutions $\mathcal{S}_\pm$ reads
    \begin{equation}
        \label{eq:Jac-pm}
        J(p_\pm,h_\pm=p_\pm/\varepsilon)=
        \begin{pmatrix}
             (1 -2\beta p_\pm )p_\pm  & (\phi -2\gamma p_\pm/\varepsilon )p_\pm\\
            1 & -\varepsilon
        \end{pmatrix},
    \end{equation}
    whose eigenvalues are
    \begin{equation}
    \label{eq:eigen-pm-O}
    \lambda_\pm(p_\sigma)= \dfrac{(1 -2\beta p_\sigma )p_\sigma-\varepsilon \pm \sqrt{\big[(1 -2\beta p_\sigma )p_\sigma-\varepsilon\big]^2+4\varepsilon p_\sigma\big[(1+\phi/\varepsilon) -2(\beta +\gamma /\varepsilon^2)p_\sigma \big]}}{2},
    \end{equation}        
    where $\sigma=\pm 1$ differentiates between the two populated states. It is easy to rewrite the previous expressions in a more useful form by making use of Eq.~(\ref{eq:fix-p-2}) to evaluate the second term in square brackets inside the square root to finally obtain 
    \begin{equation}
        \label{eq:eigen-pm}
        \lambda_\pm(p_\sigma)= \dfrac{\tau(p_\sigma) \pm \sqrt{\big(\tau(p_\sigma)\big)^2 - 4\sigma\big(\omega(p_\sigma)\big)^2 }}{2},
    \end{equation}
    with
    \begin{align}
        \tau(p_\sigma) &= (1 -2\beta p_\sigma )p_\sigma-\varepsilon, \label{eq:eigen-tau}\\
        \omega(p_\sigma) &= \sqrt{p_\sigma \sqrt{ (\phi+\varepsilon)^2 +4(1-\alpha)(\gamma+\beta\varepsilon^2)}}. \label{eq:eigen-omega}
    \end{align}

    Eq.~(\ref{eq:eigen-pm}) provides very useful information. If we look at the eigenvalues of the lowly populated solution  $\mathcal{S}_-$, by setting $\sigma=-1$ in Eq.~(\ref{eq:eigen-pm}), we see that, as long as $\mathcal{S}_-$ exists ($\alpha_T<\alpha<\alpha_{SN}$), the value of the square root is real and larger than $\tau(p_-)$. Thus, the pair of eigenvalues $\lambda_\pm$ for $\mathcal{S}_-$ are always real with opposite signs, and we conclude that $\mathcal{S}_-$ is a saddle as long as it exists. This pair of eigenvalues undergoes a saddle index change (SIC) when their quotient is equal to minus one. This occurs for $\tau(p_-)=0$, when Eq.~(\ref{eq:SIC-Condition}) is fulfilled.\\

    Now we study the stability of the highly populated solution $\mathcal{S}_+$ by setting $\sigma=+1$ in Eq.~(\ref{eq:eigen-pm}). In contrast to what occurred in the case of $\sigma=-1$, now while $\mathcal{S}_+$ exists ($\alpha<\alpha_{SN}$) the modulus of the long square root of Eq.~(\ref{eq:eigen-pm}) is smaller than $\tau(p_+)$. Thus, the sign of the pair of eigenvalues $\lambda_\pm$ is determined by the sign of $\tau(p_+)$. Therefore, the change in the stability of $\mathcal{S}_+$ occurs for $\tau(p_+)=0$, when Eq.~(\ref{eq:HopfCondition}) is fulfilled. We name $\alpha_H$ to the value of $\alpha$ at which this bifurcation occurs.\\
    
    When we substitute $\tau(p_+)=0$ into Eq.~(\ref{eq:eigen-pm}), we obtain $\lambda_\pm(p_+)= \pm i\omega(p_+)$. The expression in the most inner square root of Eq.~(\ref{eq:eigen-omega}) is always positive as long as $\alpha<\alpha_{SN}$, which is the condition for $p_+$ to exist. Since this inner square root is real (and positive) as long as $p_+$ exists (which is also positive), the argument of the long square root is positive. Hence, $\omega(p_+)$ is real and the two eigenvalues $\lambda_\pm(p_+)$ are a pure complex conjugate pair. Thus, we can conclude that when Eq.~(\ref{eq:HopfCondition}) is fulfilled, the solution $\mathcal{S}_+$ undergoes a Hopf bifurcation, from which a periodic cycle emerges. This cycle has precisely frequency $\omega$ when the amplitude of the oscillations is small. Explicit expressions for both $\alpha_{SIC}$ and $\alpha_H$ (the values at which Eq.~(\ref{eq:SIC-Condition}) and Eq.~(\ref{eq:HopfCondition}) are fulfilled, respectively) are given in Appendix~\ref{app:Explicit}.\\
    
    From now on, we will only continue the multiple scale analysis around $\mathcal{S}_+$ for $\alpha_0=\alpha_H$ since we are interested in obtaining an equation for the amplitude $A_1$ of the oscillations emerging from the Hopf bifurcation, which also allows us to determine whether they are stable or not. In order to do this, we define $J_H$, the Jacobian evaluated at $\mathcal{S}_+$ when $\alpha=\alpha_H$, which occurs when $(1-2\beta p_+)p_+=\varepsilon$. If this happens, and we express the eigenvalues of $J_H$ as $\lambda_\pm=\pm i\omega$ (as defined in Eqs.~(\ref{eq:eigen-pm}) and (\ref{eq:eigen-omega})) the following equation is fulfilled:
    \begin{equation}
        \label{eq:Jac-H-term}
        (\phi-2\gamma p_+/\varepsilon)p_+=-\omega^2-\varepsilon^2
    \end{equation}
    Thus we have
    \begin{equation}
        \label{eq:Jac-H}
        J_H=
        \begin{pmatrix}
             \varepsilon & -\omega^2-\varepsilon^2\\
            1 & -\varepsilon
        \end{pmatrix}.
    \end{equation}

    Now we can express $\mathcal{S}_1$ as a linear combination of the right eigenmodes of $J_H$, as Eq.~(\ref{eq:O-eps-1}) implies:
    \begin{equation}
        \label{eq:Ansatz-1}
        \begin{pmatrix} p_1 \\ h_1 \end{pmatrix} = \begin{pmatrix} \varepsilon + i\omega \\ 1 \end{pmatrix} A_1 e^{i\omega \tau_0} + \begin{pmatrix} \varepsilon - i\omega \\ 1 \end{pmatrix} A_1^* e^{-i\omega \tau_0}.
    \end{equation}

    We shall make some comments on this solution: The amplitude factor $A_1$ (superindex $^*$ denotes the complex conjugate) may depend on all the time scales ($\tau_1$, $\tau_2$, $\tau_3$...) except for the fast time $\tau_0$; thus having $\partial_{\tau_0} A_1 = \partial_{\tau_0} A_1^* = 0$. This is the general solution to Eq.~(\ref{eq:O-eps-1}), however, if and only if we neglect the term $-\alpha_1 p_*$. This is not a problem, since we will prove later that $\alpha_1$ has to be zero (Eq.~(\ref{eq:Solv-O2-1})).\\

    The equations for the unknown amplitude $A_1(t)$ emerge as a consequence of applying the solvability condition (Fredholm alternative) \cite{taylor1980functional} to the equations for $\mathcal{S}_2$ and $\mathcal{S}_3$ which are obtained on higher orders of $\epsilon$. For this purpose, we define a new operator
    \begin{equation}
        \label{eq:L-Operator}
        \mathcal{L} = \partial_{\tau_0} - J_H
    \end{equation}

    and the scalar product
    \begin{equation}
        \label{eq:ScalarProduct}
        \langle \mathcal{A}(\omega_1) | \mathcal{B}(\omega_2) \rangle = \dfrac{1}{T} \int_T \begin{pmatrix} a_x^* & a_y^* \end{pmatrix} \cdot \begin{pmatrix} b_x \\ b_y \end{pmatrix} e^{i(\omega_2-\omega_1)\tau_0} \, dt
    \end{equation}
    where
    \begin{equation}
        \label{eq:ScalarProduct-1}
        \mathcal{A}(\omega_1) = \begin{pmatrix} a_x \\ a_y \end{pmatrix}e^{i\omega_1\tau_0} \, \, , \, \, \mathcal{B}(\omega_2) = \begin{pmatrix} b_x \\ b_y \end{pmatrix}e^{i\omega_2\tau_0},
    \end{equation}

    being $T=2\pi/\omega$ the period associated with the fundamental frequency.\\
    
    We compute the left eigenmodes of $\mathcal{L}$
    \begin{equation}
        \label{eq:LeftEigenmodes}
         \mathcal{V} = \begin{pmatrix} 1, & -\varepsilon -i\omega \end{pmatrix} e^{i\omega \tau_0} + \begin{pmatrix} 1, & -\varepsilon +i\omega \end{pmatrix} e^{-i\omega \tau_0}
    \end{equation}
    
    since the solvability condition at each order is to ensure that the scalar product between these and the result of applying $\mathcal{L}$ to $\mathcal{S}_2$ and $\mathcal{S}_3$, respectively, is zero.\\

    At order $\epsilon^2$, we have
    \begin{equation}
    \label{eq:O-eps-2}
    \mathcal{L}\mathcal{S}_2 = \mathcal{L}\begin{pmatrix} p_2 \\ h_2 \end{pmatrix} = \Bigg[-\partial_{\tau_1} + \begin{pmatrix} -\alpha_1 & 0\\ 0 & 0 \end{pmatrix} \Bigg] \begin{pmatrix} p_1 \\ h_1 \end{pmatrix} + \begin{pmatrix}
        -\alpha_2 p_+ + (1 -3\beta p_+)p_1^2  + (\phi -2\gamma p_+/\varepsilon )p_1 h_1 - \gamma p_+ h_1^2\\
        0
    \end{pmatrix}.
    \end{equation}    
    
    The solvability condition
    \begin{equation}
        \label{eq:Solv-O2}
        \langle \mathcal{V} | \mathcal{L}\mathcal{S}_2 \rangle = 0
    \end{equation}
    implies restrictions only in the terms that are linear in $p_1$ or $h_1$ since they are the only ones including the resonant mode $e^{\pm i\omega \tau_0}$ (any other terms are orthogonal to $\mathcal{V}$ in the sense that we have defined the scalar product Eq.~(\ref{eq:ScalarProduct})). These restrictions are
    \begin{equation}
        \label{eq:Solv-O2-1}
        \alpha_1=0, \quad \quad \partial_{\tau_1}A_1 = \partial_{\tau_1}A_1^*=0.
    \end{equation}

    Hence, we propose a solution $\mathcal{S}_2$ of the form:
    \begin{equation}
        \label{eq:Ansatz-2}
        \begin{pmatrix} p_2 \\ h_2 \end{pmatrix} = \begin{pmatrix} \psi_0 \\ \xi_0 \end{pmatrix}A_0 + \begin{pmatrix} \psi_2 \\ \xi_2 \end{pmatrix} A_2 e^{2i\omega \tau_0} + \begin{pmatrix} \psi_2^* \\ \xi_2^* \end{pmatrix} A_2^* e^{-2i\omega \tau_0}.
    \end{equation}
    
    There may also be terms scaling with $e^{\pm i\omega \tau_0}$, but we can advance from the conditions given in Eq.~(\ref{eq:Solv-O2-1}) that they have to be identically zero. Again, the amplitude factor $A_2$ may depend on all time scales except the fast time $\tau_0$. Inserting this ansatz, together with Eq.~(\ref{eq:Ansatz-1}) and Eq.~(\ref{eq:Solv-O2-1}), in Eq.~(\ref{eq:O-eps-2}), we obtain:
    \begin{align}
        \psi_0 &= -\varepsilon \quad \quad \quad \quad \quad \xi_0 = -1 \label{eq:S2-1}\\
        A_0  &= \dfrac{\alpha_2 p_+}{\omega^2} +\dfrac{(\varepsilon\phi + \omega^2+\varepsilon^2)|A_1|^2}{\omega^2} \label{eq:S2-2}\\
        \psi_2  &= (\varepsilon+2i\omega) \quad \quad \xi_2 = 1 \label{eq:S2-3}\\
        A_2   &= \Bigg(\dfrac{\varepsilon\phi}{2} +\dfrac{i\omega(\omega^2-2\varepsilon^2-3i\varepsilon\omega)}{p_+} \Bigg)\dfrac{A_1^2}{3\omega^2} \label{eq:S2-4}
    \end{align}
        
    At order $\epsilon^3$, we have
    \begin{equation}
    \label{eq:O-eps-3}
        \mathcal{L}\mathcal{S}_3 = \mathcal{L}\begin{pmatrix} p_3 \\ h_3\end{pmatrix} = 
    \Bigg[-\partial_{\tau_2}+\begin{pmatrix} -\alpha_2 & 0 \\ 0 & 0 \end{pmatrix}\Bigg]\begin{pmatrix} p_1 \\ h_1 \end{pmatrix}
     + \begin{pmatrix}
         \begin{gathered}
              -\alpha_3 p_+  -\beta p_1^3 +2(1-3\beta p_+)p_1 p_2 +\\ (\phi - 2\gamma p_+/\varepsilon)(p_1 h_2 +p_2 h_1) - \gamma p_1 h_1^2 -2\gamma p_+ h_1 h_2
         \end{gathered}\\
        0
    \end{pmatrix}.
    \end{equation}

    In this equation, they also appear terms including the derivatives of $p_2$ and $h_2$ with respect to $\tau_1$. However, those terms are identically zero since we just showed that the amplitude factors $A_0$ and $A_2$ appearing in Eq.~(\ref{eq:Ansatz-2}) inherit the time dependence of $A_1$ (Eqs.~(\ref{eq:S2-2}) and (\ref{eq:S2-4})), which has zero derivative with respect to $\tau_1$ (Eq.~(\ref{eq:Solv-O2-1})).\\
    If we apply the solvability condition 
    \begin{equation}
        \label{eq:Solv-O3}
        \langle \mathcal{V} | \mathcal{L}\mathcal{S}_3 \rangle = 0,
    \end{equation}

    we obtain an equation for the time ($\tau_2$) evolution of amplitude $A_1$
    \begin{equation}
        \label{eq:dtau2-A1}
        \partial_{\tau_2} A_1 = \Phi A_1 + \Lambda |A_1|^2 A_1,
    \end{equation}

    where $\Phi$ and $\Lambda$ are complex functions that depend on all the parameters of the model. This dependence is both explicit and implicit via intermediate expressions such as $p_+$ and $\omega$:
    \begin{equation}
        \Phi = \dfrac{\alpha_2\varepsilon(p_+ - 2\varepsilon)}{2\omega^2} - i\,\dfrac{\alpha_2\varepsilon\big(2\omega^2 + (\varepsilon+\phi)p_+\big)}{2\omega^3} \label{eq:Phi}
    \end{equation}
    \begin{align}
    &\begin{aligned} \mathrm{Re}(\Lambda) &= \dfrac{\gamma p_{+}}{2 \varepsilon} + \dfrac{\big( (\phi-\gamma-3\beta\omega^2)\omega^{2} - (2\gamma+3\beta\omega^2) \phi p_{+} \big) + \big(\phi^{2} + 2 \omega^{2} + (4\gamma-3\beta\omega^{2}) p_{+}\big)\varepsilon}{2 \omega^{2}} \\
    +&\dfrac{ 3(\phi - 3\gamma -4\beta \omega^{2} - 3\beta\phi p_{+})\varepsilon^{2} + (2-3\beta p_{+})\varepsilon^{3} -9\beta\varepsilon^{4} }{2 \omega^{2}}
    \end{aligned} \label{eq:Re_Lambda} \\
    &\begin{aligned} \mathrm{Im}(\Lambda) &= \dfrac{\gamma(-\omega^2 + \phi p_{+})}{3\varepsilon\omega} + \dfrac{- (\phi^2+4\omega^2) + (6\beta\omega^2 - 19\gamma) p_{+}}{6\omega} + \dfrac{\big( (27\beta\omega^2 + 19\gamma - 14\phi)\omega^2 + (42\beta\omega^2 + 15\gamma)\phi p_{+} \big)\varepsilon }{6\omega^3} \\
    +& \dfrac{\big( -10\phi^2 - 14\omega^2 +(21\beta\omega^2 - 30\gamma) p_{+} \big)\varepsilon^2 + (72\beta\omega^2 + 45\gamma - 20\phi + 60 \beta\phi p_{+} )\varepsilon^{3} + (-10 + 15\beta p_{+})\varepsilon^4 + 45\beta\varepsilon^{5}}{6 \omega^{3}}.
    \end{aligned} \label{eq:Im_Lambda} 
    \end{align}

    Trying to make sense of these expressions seems quite a hard task. However, we do not need to do it, since we are simply looking to have an expression (Eq.~(\ref{eq:Re_Lambda})) for $\mathrm{Re}(\Lambda)$, the first Lyapunov coefficient of Eq.~(\ref{eq:dtau2-A1}) for the amplitude of oscillations emerging from the Hopf bifurcation that occurs in our system of equations. The sign of this coefficient determines whether the limit cycle is stable or not, and the locus where it is equal to zero is the Bautin bifurcation. This bifurcation divides the Hopf bifurcation line (line in the $\varepsilon$-$\alpha$ plane) into two sub-lines: a Hopf bifurcation line named $\mathrm{H^-}$ from where a stable cycle emerges and another one named $\mathrm{H^+}$ from where an unstable cycle does. A third line also emerges from this bifurcation: the fold (or saddle-node) of cycles $\mathrm{FC}$, the line in the $\varepsilon$-$\alpha$ plane in which the stable cycle emerging from $\mathrm{H^-}$ coalesces with the unstable cycle emerging from $\mathrm{SL^+}$ and both are destroyed.\\

    In order to find the Bautin bifurcation we must solve $\mathrm{Re}(\Lambda)=0$ over the Hopf line $\alpha=\alpha_H$. We first substitute $\omega^2$ by one of its correspondent expressions in terms of the model parameters and $p_+$ in Eq.~(\ref{eq:Re_Lambda}). The one that better suits our purpose now is $\omega^2 = -\varepsilon^2 - (\phi-2\gamma p_+/\varepsilon)p_+$ (Eq.~(\ref{eq:Jac-H-term})). Then, we indirectly set $\alpha=\alpha_H$ by substituting $p_+$ by its correspondent expression when the Hopf bifurcation occurs, which is the solution to Eq.~(\ref{eq:HopfCondition}), given by the positive branch of Eq.~(\ref{eq:p_SIC-Hopf}). The resulting expression is a 25th-degree polynomial in $\varepsilon$ that can only be solved numerically. For the set of parameters used in this work ($\phi=1$, $\beta=\gamma=0.25$), the solution we are looking for is $\varepsilon_B = 0.464101(6)$ and its corresponding value of $\alpha_B = 2.732050(8)$, obtained by inserting $\varepsilon_B$ in the expression of $\alpha_H$ in terms of the rest of the parameters of the model, which is the positive branch of Eq.~(\ref{eq:alfa_SIC-Hopf}).\\

    Finally, it should be noted that this model does not exhibit Turing instabilities that could offer alternative explanations to coral reef pattern formation. When examining the linear stability of the homogeneous solutions $\mathcal{S}_0$ and $\mathcal{S}_\pm$ under periodic perturbations of wavenumber $q$, it is observed that the eigenvalues remain largely unchanged. Specifically, for the bare state $\mathcal{S}_0$, the first eigenvalue $\lambda_\alpha=1-\alpha$ adjusts to $\lambda_{\alpha,q}=1-\alpha-q^2$. For the populated states $\mathcal{S}_\pm$, the eigenvalues $\lambda_\pm(p_\sigma)$ given by Eq.~(\ref{eq:eigen-pm}) are only altered through $\tau(p_\sigma)$ according to Eq.~(\ref{eq:eigen-tau}), which now becomes $\tau_q(p_\sigma) = \tau(p_\sigma) - q^2$ and the term $\big(\omega(p_\sigma)\big)^2$, which now becomes $\big(\omega_q(p_\sigma)\big)^2=\big(\omega(p_\sigma)\big)^2 + \sigma\varepsilon q^2$. In all instances, the addition of the $q^2$ term does not introduce new instabilities. Homogeneous solutions display greater stability against periodic perturbations compared to homogeneous ones, as the real part of the eigenvalues that could swift from negative to positive are globally reduced by the inclusion of the $-q^2$ term.

    \section{\label{app:Explicit}Explicit expressions for some lines in parameter space\protect}

    We can derive explicit expressions for bifurcation lines that are challenging to handle through analytic calculations but prove useful in numerical computations.\\
    First, we invert Eq.~(\ref{eq:fix-p-2}) for the values of populated stationary solutions $p_\pm$ as a function of $\alpha$ in order to obtain this bifurcation parameter in terms of the stationary population $p_*$:
    \begin{equation}
        \label{eq:alfa_p}
       \alpha = 1 + \dfrac{(\phi+\varepsilon)^2 - \Big[ 2(\gamma+\beta\varepsilon^2)p_*/\varepsilon - (\phi+\varepsilon) \Big]^2}{4(\gamma+\beta\varepsilon^2)},
    \end{equation}

    Second, we solve both Eq.~(\ref{eq:SIC-Condition}) and Eq.~(\ref{eq:HopfCondition}) (they are essentially the same equation) for $p_*$
    \begin{equation}
        \label{eq:p_SIC-Hopf}
        p_{*} = \dfrac{1\pm\sqrt{1-8\beta\varepsilon}}{4\beta},
    \end{equation}

    and insert this expression in Eq.~(\ref{eq:alfa_p}) to obtain
    \begin{equation}
        \label{eq:alfa_SIC-Hopf}
       \alpha_{\pm} = 1 + \dfrac{(\phi+\varepsilon)^2 - \Big[\dfrac{(\gamma+\beta\varepsilon^2)(1\pm\sqrt{1-8\beta\varepsilon})}{2\beta\varepsilon}  - (\phi+\varepsilon) \Big]^2}{4(\gamma+\beta\varepsilon^2)}.
    \end{equation}
    
    Note that this equation does not distinguish explicitly between the SIC and the Hopf bifurcation. It has two branches differentiated by the sign of the inner square root that meet at $8\beta\varepsilon=1$, but these branches do not have a univocal correspondence with the lines we are interested in. This occurs because Eq.~(\ref{eq:alfa_p}) can show the same value for $\alpha$ given different values of $p_*$. It is easy, however, to select what regions defined by Eq.~(\ref{eq:alfa_SIC-Hopf}) correspond to what bifurcation by both invoking topological restrictions on the position of bifurcation lines in Fig.~\ref{fig:Bifurcations-Zoom} and also by solving numerically Eqs.~(\ref{eq:SIC-Condition}) and (\ref{eq:HopfCondition}).\\
    The Hopf bifurcation line is given by the positive branch of Eq.~(\ref{eq:alfa_SIC-Hopf}) for values of $\varepsilon$ less or equal than $\varepsilon_{TB}$ (the value of $\varepsilon$ at which the Takens-Bogdanov bifurcation occurs). The entire negative branch of Eq.~(\ref{eq:alfa_SIC-Hopf}) together with the part of the positive branch in the region $\varepsilon_{TB}\leq \varepsilon \leq 1/8\beta$ corresponds to the line where the SIC occurs.\\

    \section{\label{app:Methods}Methods\protect}

    Several numeric techniques have been used throughout the elaboration of this paper. Besides the direct evaluation of analytic expressions such as those we obtained for some bifurcations lines such as the Transcritical ($\alpha=1$), the Saddle Node (Eq.~(\ref{eq:alpha-SN})) and Hopf or SIC lines (Eq.~(\ref{eq:alfa_SIC-Hopf})); we have made extensive use of the Newton-Raphson (NR) algorithm in order to obtain solutions for several non-linear equations in this work. The simplest cases for the application of this technique are the computation of both the values for $\alpha$ and $\varepsilon$ at the Takens-Bogdanov and Bautin points, following the procedures to obtain their correspondent non-linear equations explained in Sec.~\ref{sec:Global Bifurcations} and App.~\ref{app:Multiscale analysis}, respectively. The location of these two points allows us to split the Hopf and Saddle Node lines into their respective stable and unstable branches. As mentioned in the main text, we have fixed the less relevant parameters of the model to the values $\phi=1$, $\beta=\gamma=1/4$ and $\eta=0$ in this work.\\
    
    More elaborated implementations of the NR algorithm were used when looking for other bifurcation lines, such as the Saddle Loop. In this bifurcation, homogeneous solutions following limit cycles suffer a divergence in their period as they become an homoclinic orbit. Hence, in order to track this bifurcation, one can solve Eqs.~(\ref{eq:dtp}-\ref{eq:dth}) for homogeneous solutions over a single cycle. If this cycle has period $T$, one can rescale the temporal axis defining $\tau=t/T$ and thus, our problem now is to solve
    \begin{align}
            \partial_\tau p - T\Big[ (1-\alpha)+ p -\beta p^2 +\phi h -\gamma h^2  \Big]p &= 0,\label{eq:dtp-tau} \\
        \mathrlap{}
        \partial_\tau h - T(p - \varepsilon h) &= 0, \label{eq:dth-tau}
    \end{align}
    with periodic boundary conditions $p(0)=p(1)$ and $h(0)=h(1)$.\\
    We can take advantage of this last requisite and evaluate the time derivatives making use of Fourier analysis. The evaluation of derivatives when we discretize our integration domain results in linear combinations of the variables evaluated at different points. For this reason, all the derivatives which conform the Jacobian matrix for the NR method are well defined and we can proceed using this methodology.\\
    We have introduced, however, a new variable $T$ that requires a new equation. The traditional approach is to solve, simultaneously with Eqs.~(\ref{eq:dtp-tau}) and (\ref{eq:dth-tau}) the so-called Integral Phase Condition (IPC) \cite{Beyn2007} which not only makes the whole system of equations solvable but provides a way of fixing the phase of the cycle. Our election for the IPC is:
    \begin{equation}
        \label{eq:IPC}
        \mathcal{I}\big(p_{k}(\tau),h_{k}(\tau)\big) = \int_0^1 p_{k}(\tau)\cdot\partial_\tau p_{k-1}(\tau) + h_{k}(\tau)\cdot\partial_\tau h_{k-1}(\tau) \, d\tau = 0,
    \end{equation}
    where $k$ indexes different solutions of these equations as we change the bifurcation parameters.\\
    One can fix the parameters of the model and solve Eqs.~(\ref{eq:dtp-tau}-\ref{eq:IPC}) using Newton-Raphson method in order to obtain the values of $p$, $h$ and $T$ or one can do the trick of fixing every parameter except one, that will be $\alpha$ in our case, and fix the cycle period $T$. Recalling the fact that the cycle period diverges at the SL bifurcation, we can take an arbitrarily large value of the period, which in our case is $T_\mathrm{SL}=60$, and consider that such a large period properly represents the bifurcation. Then, the NR algorithm will provide the values of $p$, $h$ and $\alpha_\mathrm{SL}$. Repeating the process for various values of $\varepsilon$ one can reconstruct the $\mathrm{SL}$ line we show in the main text. We checked at some points scattered across this computed line that increasing the selected value  of $T_\mathrm{SL}=60$ by several tens results in changes to the third significant figure of $\alpha_\mathrm{SL}$, being this level of precision is more than sufficient for the scope of our study. We have discretised the time domain with 4096 points and used a tolerance for the NR algorithm of $10^{-10}$.\\
    With these results, the Resonant Side Switching point has been computed as the point in which the linear interpolations of the points defining both the SL and SIC lines cross each other. This point allows us to split the SL line into its respective stable and unstable branches.\\

    The same technique can be used to solve the equations that determine the shape of the traveling pulses (TPs). TPs can be analyzed using a single variable \( \xi = x - ct \), where the sign of  \( c \) indicates the direction of travel, to the right if positive and to the left if negative.  \( \xi \) is then the spatial coordinate in a reference frame moving with the pulse. In this reference frame, $p(x,t)=p(\xi)$ and $h(x,t)=h(\xi)$ and the original system of two partial differential equations Eqs.~(\ref{eq:dtp}-\ref{eq:dth}) can be recast to a system of three ordinary differential equations:
     \begin{align}
        \partial_\xi p = & \, p_\xi \label{eq:dxip}\\
        \begin{split}
            \partial_\xi p_\xi = &-\Big[ (1-\alpha)+ p -\beta p^2 +\phi h -\gamma h^2  \Big]p \\
            &-\eta p_\xi^2 -c p_\xi, \label{eq:d2xip}
        \end{split} \\
        \mathrlap{}
        \partial_\xi h = &-c^{-1}(p - \varepsilon h). \label{eq:dxih}
    \end{align}
    
    For the numerical implementation, Eq.~(\ref{eq:dxip}) is redundant, and we shall rewrite Eq.~(\ref{eq:d2xip}) as
    \begin{equation}
        \partial_\xi^2 p + c  \partial_\xi p +\eta ( \partial_\xi p)^2 +\Big[ (1-\alpha)+ p -\beta p^2 +\phi h -\gamma h^2  \Big]p= 0. \label{eq:d2xip-1}
    \end{equation}
        
    As it happened before with the cycle period $T$, the pulse velocity $c$ also requires an additional equation for the system of equations to be solvable. The equation for it is again the IPC (Eq.~(\ref{eq:IPC})). In this case, there is no rescaling of the independent variable $\xi$, but the periodic boundary conditions remain. We have to ensure that the integration domain is large enough for the pulse to fit inside it, which in our case results in a domain length of $L=1000$. Again, we discretise the time domain with 4096 points and use a tolerance for the NR algorithm of $10^{-10}$.\\
    For a proper continuation of the TP when varying the value of $\alpha$, we have additionally implemented a Keller's pseudoarclength continuation algorithm. For more details about this methodology, check \cite{Beyn2007}.\\
    We have run the continuation algorithm for various values of $\varepsilon$, and we have determined the position of the SN-TP bifurcation as the line of points for which the sign in the changes in $\alpha$ swaps from positive to negative.\\
    In order to determine the position of the $\mathrm{T}_1$ and $\mathrm{T}_2$ lines, we have taken advantage of the fact that near the T-point the pulses grow large tails, diverging their size exactly at the bifurcation. The size of those tails scales with the logarithm of the distance between the actual value of $\alpha$ and the value at which the bifurcation occurs \cite{Arinyo-i-Prats2021}. This produces very little change in the value of $\alpha$ as we approach the bifurcation with a continuation algorithm. Hence, we have explored the different solutions for Eqs.~(\ref{eq:dxih}) and (\ref{eq:d2xip-1}) and when the changes in $\alpha$ given by the continuation algorithm drop below $10^{-6}$ (in absolute value), we assume that we are sufficiently close to the actual bifurcations. To reconstruct the entire lines, we simply repeat this process for several values of $\varepsilon$.\\
    With these three lines related to TPs computed, the $\shin$-point has been computed as the nearest point between the linear extrapolations of lines $\mathrm{T}_1$, $\mathrm{T}_2$ and Saddle Node of traveling Pulses.\\

    Finally, we have integrated several time trajectories of Eqs.~(\ref{eq:dtp}) and (\ref{eq:dth}) to both be used as initial guesses for the NR algorithm and also to determine the two remaining lines: The Fold of Cycles (FC) and the Spatial Excitability Limit (SEL). We integrate the trajectories using a Pseudo-Spectral Runge-Kutta 4 scheme \cite{Pere-Raul} with a time step of $\Delta t=0.05$, domain length of $L=1000$ and 8192 points for the domain discretisation.\\
    In order to determine the FC line, we have run homogeneous trajectories initialising the system with values of $\alpha$ and $\varepsilon$ below the Hopf line letting the system arrive to a stable cycle (if there is any) within 100 time units. If a cycle is found, we increase the value of $\alpha$ by 0.1 and let the system evolve again for another 100 time units. We repeat this process until the cycle disappears. When this happens, we return to the last value of $\alpha$ for which there was a stable limit cycle and we proceed to increase the value of $\alpha$ by 0.01. We repeat the same process as we did before and when the cycle disappears again, we set the accuracy to 0.001. For our purpose, this precision in $\alpha$ is more than enough and we stop the algorithm the last time the cycle vanishes, setting the value of $\alpha_\mathrm{FC}$ as the midpoint between the last two values of $\alpha$ the algorithm has used. We reconstruct the whole FC line repeating this process for values of $\varepsilon$ between $\varepsilon_\mathrm{RSS}$ and $\varepsilon_\mathrm{B}$.\\

    This idea of using nested intervals of precision to determine the region of stability of some kind of solution (stable limit cycles in the paragraph before) is extended also to examine the stability of spatially heterogeneous solutions. In this case, we set an initial condition given by a supergaussian with a plateau a little above the saddle point
    \begin{equation}
        \label{eq:Supergaussian}
        \begin{pmatrix}
            p(x,0) \\ h(x,0)
        \end{pmatrix} =  1.2 \cdot h_-
        \begin{pmatrix}
            \varepsilon \\ 1
        \end{pmatrix} \exp\Bigg ( -\Big(\dfrac{x-L/2}{2\sigma^2}\Big)^4\Bigg ),
    \end{equation}
    where $\sigma=128\cdot L/8191$ in these simulations. With $L=1000$, we have $\sigma\approx 15.63$.\\
    The SEL is calculated as the line that separates the region in which this initial condition decays to zero after 200 time units from the region in which this initial condition evolves following self-sustained dynamics. To achieve a precision of 0.001 in $\alpha$, we explore the parameter space with the nested intervals algorithm as explained before.\\

\end{document}